\documentclass{aa}  

\usepackage{graphicx}
\usepackage{txfonts}
\usepackage{graphicx}	
\usepackage{amsmath}	
\usepackage{amssymb}	
\usepackage{tikz}
\usepackage{amsmath}
\usepackage{siunitx}
\usepackage{multirow}
\usepackage{natbib}
\usepackage{cases}
\usepackage{multicol}
\usepackage{threeparttable}
\usepackage{hyperref}
\usepackage{placeins}

\newcommand{\daoa}{{\ensuremath \Delta\alpha/\alpha}}
\newcommand{\kmps}[1]{\SI{#1}{\kilo\meter\per\second}}
\newcommand{\mps}[1]{\SI{#1}{\meter\per\second}}
\newcommand{\nm}[1]{\SI{#1}{\nano\meter}}
\newcommand{\cmps}[1]{\SI{#1}{\centi\meter\per\second}}
\renewcommand{\ion}[2]{#1\,{\sc #2}}
\newcommand{\iodine}{I$_2$}

\newcommand{\secref}[1]{Sec.\,\ref{#1}}
\newcommand{\figref}[1]{Fig.\,\ref{#1}}
\newcommand{\tabref}[1]{Tab.\,\ref{#1}}

\newcommand{\ts}{\textsuperscript}
\newcommand{\ak}{A\&K}

\DeclareSIUnit\pixel{\text {pix}}
\newcommand{\pix}[1]{\SI{#1}{\pixel}}
\DeclareSIUnit{\Cnm}{\,\SI{100}{\nano\meter}}

\makeatletter
\renewcommand*\aa@pageof{, page \thepage{} of \pageref*{LastPage}}
\makeatother

\begin{document}

   \title{A new method for instrumental profile reconstruction of high-resolution spectrographs\thanks{The HARPS instrumental profile produced in this work is publicly available at \url{https://zenodo.org/doi/10.5281/zenodo.10492989} \citep{Milakovic2024_HARPS_IP_dataset}. The same data can be retrieved from the CDS via anonymous ftp to cdsarc.u-strasbg.fr (130.79.128.5) or via \url{http://cdsweb.u-strasbg.fr/cgi-bin/qcat?J/A+A/.}} \thanks{Based on observations collected at the European Southern Observatory under ESO programme 0102.A-0697(A).}}


   \author{D. Milakovi{\'c}
          \inst{1,2}
          \and
          P. Jethwa\inst{3}
          }

   \institute{
             Institute for Fundamental Physics of the Universe, Via Beirut, 2, 34151 Trieste, Italy 
        \and 
            INAF, Osservatorio Astronomico di Trieste, via Tiepolo 11, 34131, Trieste, Italy\\
              \email{dinko@milakovic.net}
         \and
             University of Vienna, Department of Astrophysics, T{\"u}rkenschanzstra{\ss}e 17, A-1180 Vienna, Austria
             }

   \date{Received 09 Nov 2023 ; accepted 18 Jan 2024}

 
  \abstract
   {Knowledge of the spectrograph's instrumental profile (IP) provides important information needed for wavelength calibration and for the use in scientific analyses. }
    {
    This work develops new methods for IP reconstruction in high-resolution spectrographs equipped with astronomical laser frequency comb (astrocomb) calibration systems and assesses the impact that assumptions on the IP shape have on achieving accurate spectroscopic measurements. 
    }
   {
   Astrocombs produce $\approx10000$ bright, unresolved emission lines with known wavelengths, making them excellent probes of the IP. New methods based on Gaussian process regression were developed to extract detailed information on the IP shape from these data. Applying them to HARPS, an extremely stable spectrograph installed on the ESO 3.6m telescope, we reconstructed its IP at 512 locations of the detector, covering 60\% of the total detector area. 
   }
   {
   We found that the HARPS IP is asymmetric and that it varies smoothly across the detector. Empirical IP models provide a wavelength accuracy better than \mps{10} (\mps{5}) with a 92\% (64\%) probability. In comparison, reaching the same accuracy has a probability of only 29\% (8\%) when a Gaussian IP shape is assumed. Furthermore, the Gaussian assumption is associated with intra-order and inter-order distortions in the HARPS wavelength scale as large as \mps{60}. The spatial distribution of these distortions suggests they may be related to spectrograph optics and therefore may generally appear in cross-dispersed echelle spectrographs when Gaussian IPs are used. Empirical IP models are provided as supplementary material in machine readable format. We also provide a method to correct the distortions in astrocomb calibrations made under the Gaussian IP assumption.
   }
   {Methods presented here can be applied to other instruments equipped with astrocombs, such as ESPRESSO, but also ANDES and G-CLEF in the future. The empirical IPs are crucial for obtaining objective and unbiased measurements of fundamental constants from high-resolution spectra, as well as measurements of the redshift drift, isotopic abundances, and other science cases.}


   \maketitle
%

\section{Introduction}
Properties of the instrumental profile (IP) are important in several aspects when making accurate spectroscopic measurements. Accurately measuring centres of spectral features is known to crucially depend on assumptions made on the IP shape, as inaccurate assumptions unavoidably bias centre measurements. When used to measure the centres of features used for instrument calibration, this inevitably leads to distortions in the wavelength calibration of the instrument. Avoiding distortions is important for fine structure constant measurements \citep{Rahmani2013,Molaro2013A&A...555A..68M,Evans2014,Whitmore2015,Dumont2017} and other science cases. Furthermore, accurate IP shape models are also needed whenever theoretical models must be compared against the data during spectral modelling. 
\begin{figure*}
    \centering
    \includegraphics[width=\textwidth]{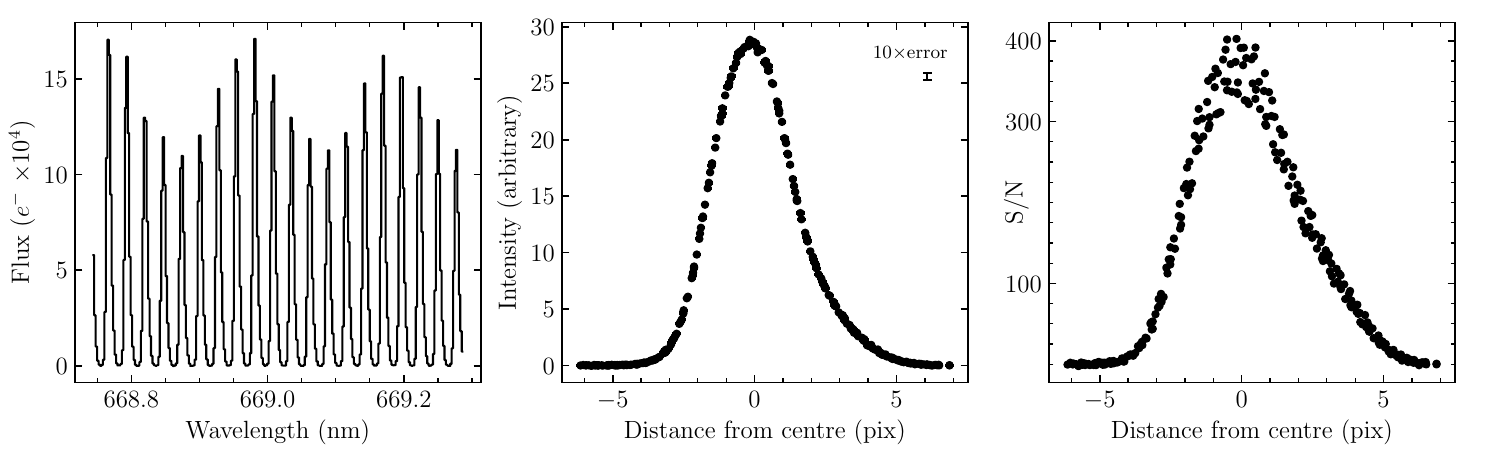}
    \caption{Basic principles of IP reconstruction methods used in this work. Left panel: The extracted astrocomb spectrum observed by HARPS in the wavelength range $\SI{668.8}{\nano\meter}\leqslant\lambda\leqslant\SI{669.3}{\nano\meter}$ containing 20 astrocomb modes, blaze-corrected and background-subtracted. Middle panel: The same lines as in the left panel, flux normalised and stacked on their best-fit Gaussian centres (see \secref{sec:fitting_IP}). The result is a remarkably well-sampled representation of the 1d IP in this wavelength range with 256 points. The sub-pixel sampling was achieved due to the IP falling slightly differently with respect to the pixel centre for each line. The right-skewed asymmetry in the IP shape is immediately noticeable. The error bar in the top right corner shows ten times the median error on individual points (enlarged for visibility). Right panel: Individual points reach an S/N of several hundred across a large fraction of the HARPS wavelength range.}
    \label{fig:stack_illustration}
\end{figure*}

Most spectrographs have an approximately Gaussian IP, with strong departures from this shape generally caused by optical design choices or (rarely) defects in the spectrograph construction. The IP shape has received comparatively little attention so far, for several reasons. Firstly, before the advent of extremely precise spectrographs, spectrograph performance was limited by systematic effects other than the IP shape. Secondly, in radial velocity studies, the exact assumptions of IP shape should be of little consequence as long as the shape is applied consistently, especially when comparing multi-epoch observations in which spectral features appear at approximately the same locations on the detector. Lastly, the ideal way for IP shape reconstruction is by observing a set of unresolved, uniformly distributed, and unblendend spectral features covering a large fraction of the instrument's spectral range and such sources just recently became available.

The IP for several instruments was previously reconstructed from spectra in which an absorption cell containing molecular iodine (\iodine) was inserted in the light path while observing astronomical objects \citep{Marcy1992,Valenti1995,Kambe2002}. The \iodine~cell methods have the benefit of using the light following the same optical path as the scientific observations so the reconstructed IP is exactly the one that appears in the data. However, the strong absorption from \iodine~makes it unfeasible to use when observing faint targets. Another drawback comes from the fact that the lines in the \iodine~spectrum are blended even when observed at a spectral resolution of $\sim 10^{6}$, complicating IP reconstruction. A comprehensive description on the use of the \iodine~cell for IP reconstruction can be found in \citet{Butler1996}. 

A different approach is offered by the astronomical laser frequency comb (LFC) systems or `astrocombs'. Astrocomb lines have almost all necessary properties to make them suitable tools for IP reconstruction, as we explain in \secref{sec:astrocomb_as_a_tool}. Several other groups have recently used an astrocomb for this purpose \cite{Zhao2019,Hirano2020}.

The paper is organised as follows. \secref{sec:instrumental_profile} presents the theoretical considerations of our work. New methods presented in \secref{sec:methods}, based on Gaussian process (GP) regression and iterative procedures, allow for the reconstruction of the most likely non-parametric function describing the IP. These methods were applied to reconstruct the IP of HARPS using its astrocomb system in \ref{sec:IP_reconstruction}. Improvements from knowing the empirical IP are presented in \secref{sec:calibration_results}. The main results are summarised in \secref{sec:result_summary} followed by a discussion in \secref{sec:discussion}. We conclude with a discussion of future prospects in \secref{sec:conclusions}. 

\section{The instrumental profile of a spectrograph}\label{sec:instrumental_profile}
\subsection{Astrocomb as a tool for IP reconstruction}\label{sec:astrocomb_as_a_tool}
The LFC \citep{Udem2002,Haensch2006} produces a train of emission lines (also known as modes) with frequencies determined by the equation
\begin{equation}\label{eq:lfc}
    f_n = f_0 + n f_r,
\end{equation}
where $f_n$ is the frequency of the $n^{\rm th}$ mode and $f_0$ and $f_r$ are the offset and repetition frequencies. Both $f_0$ and $f_r$ are actively controlled frequencies in the MHz range, stabilised against a radio-frequency standard such as an atomic clock \citep[see, e.g., ][for more details]{Probst2015PhD}. As such, the LFC line frequencies are known with a relative accuracy of $\approx 10^{-12}$ as long as a frequency lock with the reference is maintained. The accuracy and stability of the frequency standards can be transferred to the spectrograph, making LFCs the ideal wavelength calibration for astronomical spectrographs. 

LFCs specifically adapted for astronomical spectrograph calibration are named astrocombs and have $f_n$ and $f_r$ in the GHz range to match the spectrograph resolving power. First astrocomb prototypes were tested on the German Vacuum Tower Telescope \citep{Steinmetz2008} and the HARPS  \citep{Wilken2010} spectrographs. The left panel of \figref{fig:stack_illustration} shows a small part of the astrocomb spectrum observed by HARPS. 

In the context of IP reconstruction considered here, astrocombs show another useful property: the comb lines are intrinsically extremely narrow,  \SIrange{300}{500}{\kilo\hertz} (Tilo Steinmetz, private communication). This corresponds to a relative uncertainty of \SIrange{5e-10}{1e-9}{} (or  $\lesssim\mps{1}$, when scaled by the speed of light). For comparison, the resolution element width of a spectrograph with spectral resolving power $R=\lambda/\delta\lambda=f/ \delta f = 120\,000$ (where $\delta\lambda$ and $\delta f$ are the full width at half maximum, FWHM, of the resolution element in wavelength and frequency space, respectively) at \SI{550}{\nano\meter} is $\delta f = \SI{4.54}{\giga\hertz}$. This is 10\,000 times broader than the intrinsic astrocomb line width. The astrocomb lines are therefore completely unresolved by even the highest resolution astronomical spectrographs and, as such, individual modes can be considered monochromatic impulse inputs to the optical system. Their image on the detector is therefore a direct digitised representation of the 2d IP. Correspondingly, the extracted astrocomb spectrum then represents the 1d IP. For example, the middle panel of \figref{fig:stack_illustration} shows a stack of flux normalised astrocomb lines observed by HARPS on top of their best-fit Gaussian centres.

Lastly, the astrocomb lines are high signal-to-noise (S/N), as seen in the right panel of \figref{fig:stack_illustration}. They also have very uniform flux levels: the dynamic range of the HARPS astrocomb peak amplitudes is $\approx\SI{4}{\dB}$. We are therefore able to use a single astrocomb calibration frame to reconstruct the IP without needing to worry about its variation with recorded light intensity within a single frame. In fact, exposures taken with different flux levels can be used to determine the relation between light intensity and IP shape in the future. 

\subsection{The digitised IP}\label{sec:effective_IP}

This work was heavily influenced by the paper by \cite{Anderson2000}, henceforth referred to as \ak. While they were interested in performing accurate astrometric measurements and hence the 2-dimensional IP and we are interested in accurate wavelength measurements and hence the 1-dimensional IP, most of the considerations made by the authors apply here unchanged. 

Similarly to \cite{Valenti1995} previously, \ak~considered the fact that a pixelated detector does not directly record information on the `true' instrumental profile (their `iPSF'). Instead, what is recorded is the integrated product of the iPSF with the 2-dimensional pixel sensitivity profile ($\mathcal{R}$), where the integration limits are determined by the pixel's boundaries (see their section 3 and formulae (1)-(5), specifically). This led them to define the `effective' IP (their `ePSF') as the convolution of the iPSF with $\mathcal{R}$. The digitised image of a point source on the detector then provides samples of ePSF, with the values in individual pixels determined precisely by the distance of the centre of the pixel from the point source, multiplied by some flux factor intrinsic to the source. 

Importantly, \ak~showed that one does not need to determine the iPSF to measure the positions of features with subpixel accuracy. Instead, determining ePSF is sufficient because the recorded pixel values -- the only observable -- are already integrated over the pixel area with appropriate weights determined by the pixel response function. Going further, \ak~demonstrated how repeated observations of the same point sources, each offset from the other by $\lesssim \pix{1}$, can be used to reconstruct ePSF on a subpixel scale and its variation across the detector. 

Their approach has three major advantages, as noted by the authors themselves. Firstly, it avoids integrating the true IP over the pixel area when fitting for the position of the feature (the centre of a star image in their case, and the centre of an astrocomb line for us), as the integration over the area of the pixel is already included into the definition of the effective IP. Secondly, fewer computational resources are needed to solve for the effective IP than the true IP itself. Finally, because of the definition of the ePSF as the convolution of iPSF and $\mathcal{R}$, the latter is automatically incorporated into the modelling process and needs not to be untangled from the recorded point source image. 

We agreed with the reasoning of \ak~above, and have applied it to reconstruct the 1-dimensional IP. All references to the IP in the rest of the paper therefore refer to the effective IP, unless stated otherwise.

\subsection{The 1-dimensional effective IP}\label{sec:1d_effective_IP}
The flux in an extracted spectral pixel is given by:
\begin{equation}\label{eq:flux_i}
    F_{i} = f_* \int_{-\infty}^{+\infty} \psi_I (x-x_*) \mathcal{Q} (x-i)\, {\rm d}x + B_{i},
\end{equation}
where the index $i$ identifies the pixel centred at $x=i$ (in a single echelle order). The values $f_*$ and $x_*$ are the total integrated flux (brightness) and the position of a monochromatic light source, respectively, and $B_{i}$ is the background level. $\psi_I(\Delta x)$ is the 1-dimensional true IP, that is it quantifies the fraction of light (per 1d pixel unit length) that falls onto the point offset by $\Delta x$ from $x_*$. Equation \eqref{eq:flux_i} is analogous to \ak 's equation (1) and our $\psi_I$ is a 1d analogue of their iPSF.

$\mathcal{Q}$ replaces \ak 's $\mathcal{R}$ in our equations. In addition to the 2d pixel response function contained in the latter, $\mathcal{Q}$ carries the information on all other operations applied to the data by the data acquisition and spectral extraction processes. This includes bias subtraction, flat-fielding, weighted summation of pixels in the spatial direction, etc. Untangling contributions of each of these operations in 1d spectra is impossible in practice, and doing so would require working on 2d raw data. However, like with $\mathcal{R}$, it is not necessary to know $\mathcal{Q}$'s value.

Applying the same transformations as in \ak , appropriately modified for 1d, we obtain:
\begin{equation}\label{eq:pixel_flux}
    F_{i} = f_* \psi_E \left(i-x_*\right) + B_i,
\end{equation}
where:
\begin{equation}\label{eq:effective_ip_definition}
    \psi_E (\Delta x) = \int_{-\infty}^{+\infty} \mathcal{Q}'(x) \psi_I (\Delta x - x)\, {\rm d}x ,
\end{equation}
and $\mathcal{Q}'(x)=\mathcal{Q}(-x)$. Equation \eqref{eq:effective_ip_definition} is the 1d analogue of \ak 's ePSF. %

In above Equations, $x$ stands for detector pixels in uncalibrated data but the same reasoning and Equations would be valid if $x$ were replaced by $\lambda$ in wavelength calibrated data. In fact, we do this later, deriving two sets of IP shapes. The first set was derived from non-calibrated data and used to determine the centres of astrocomb lines in pixels. These centres were used to wavelength calibrate the instrument, allowing for a second set of IP shapes to be produced from the same data using $\lambda$. The latter were used to investigate the accuracy of the wavelength calibration in \secref{sec:calibration_results}.

\subsection{GP: non-parametric functional form for the IP}\label{sec:GP_regression}
Previous studies modelled the non-Gaussian shapes using different modifications to the Gaussian, including modifying the wings \citep{Cardelli1990} or adding satellite Gaussian profiles \citep{Valenti1995,Zhao2019}. More recently, a Gaussian with a modified exponent in combination with cubic splines to modify that shape when necessary \citep{Hao2020}. \ak~used splines to describe the full IP shape. While we tried out some of these approaches, we were unable to obtain satisfying results. Ultimately, we decided to use GP regression to obtain a smooth function that is most likely representation of the HARPS instrumental profile. This is similar to what \citet{Hirano2020} have done for the InfraRed Doppler spectrograph on the Subaru telescope.

GP regression is a technique for probabilistic, non-parametric regression \citep{RasmussenWilliams06}. Here we provide a basic overview; see \citet{AigrainForemanMackey22} for a detailed review in an astronomical context. Given pairs of data-points $(x_i,y_i)$ with noise $\epsilon_i$, the goal of regression is to find a function $f$ such that $y_i=f(x_i)+\epsilon_i$. A GP is a prior over this space of functions. The defining property of a GP is that for any finite sample of points $\mathbf{x}$, the sampled function values $\mathbf{f}=\{f(\mathbf{x})\}$ have a multivariate normal probability distribution
\begin{equation}\label{eq:gp_probability_distribution}
    p(\mathbf{f}) = \mathcal{N} \left(\mathbf{m}, \mathbf{K} \right),
\end{equation} 
which is parameterised by a mean vector $\mathbf{m}$ and covariance matrix $\mathbf{K}$. These are, in turn, specified by a mean function $m$ and covariance function $k$, where
\begin{align}\label{eq:gp_definitions}
    &m_i = m(x_i;\theta), \\
    &K_{ij} = k(x_i, x_j;\phi),
\end{align}
which depend on hyperparameters $\theta$ and $\phi$ respectively. Assuming that the noise $\epsilon_i$ is normally distributed with zero mean and scale $\sigma_i$, i.e. 
\begin{equation}
    p(\epsilon_i) = \mathcal{N}(0, \sigma_i),
\end{equation} 
then the likelihood of the data $\mathbf{y}$ as a function of the hyperparameters is given by  
\begin{equation}\label{eq:likelihood}
    \mathcal{L}(\theta,\psi) = \mathcal{N}(\mathbf{y} ; \mathbf{m}, \mathbf{K}+\mathrm{diag}(\{\sigma_i\})).
\end{equation}

Given observed data $(x_i,y_i)$, training the GP means to find the hyperparameters $(\theta,\phi)$ which maximise this likelihood. Once trained, the GP encodes a probability distribution over suitable regression functions $f$. Ensembles of functions $f$ can be sampled from the trained GP (see \citet{AigrainForemanMackey22} for details), and the variance amongst these functions encodes the uncertainty in the regression. The specific choices of mean functions and covariance functions that we adopt in this work are introduced in later sections: in section \ref{sec:GP_for_IP} we describe a GP model for the line profile, and in section \ref{sec:empirical_variance}, a GP to describe the observational errors.

%
\section{Methods}\label{sec:methods}
%

The instrumental profile can be estimated from the astrocomb spectra by inverting Eq.~\eqref{eq:pixel_flux}:
\begin{equation}\label{eq:effective_IP_estimator}
    \hat{\psi} (\Delta x) = \frac{F_i-B_i} {f_*},
\end{equation}
where we have now dropped the subscript ``$E$'' on $\psi_E$ for the ease of notation. The hat above $\psi$ indicates estimation from observations. In the equation above, $\Delta x=i-x_*$ is the distance between the pixel and astrocomb line centres and $f_*$ is the integrated astrocomb line flux. Knowing $x_*$ and $f_*$ allows for determining at which offset $\Delta x$ has the pixel centered at $x=i$ sampled $\psi$ and how the sampling has been scaled. Stacking the observations on top of each other provides subpixel sampling of $\psi$, as shown in the middle panel of \figref{fig:stack_illustration}. The implicit assumption in performing the stacking is that $\psi$ does not change across the range covered by the lines being stacked, but we relax this assumption later by allowing $\psi$ to be different at the location of each astrocomb line. 

When performing the GP regression, we considered also the variance on $\hat{\psi}$. It is, however, not simple to calculate this quantity analytically because it means calculating the variance of the ratio of two random Poisson variables (the first one being $F_i - B_i$ and second one being $f_*$), one of which is conditional on the other, that is $f_*=\sum_i(F_i-B_i$). Instead, we used the following simple approximation that preserves S/N in the data:
\begin{equation}\label{eq:effective_IP_variance}
    \sigma^2_{\hat{\psi}} = \frac{1}{f_*^2} \left[\sigma_{F_i}^2 + \sigma_{B_i}^2 \right] .
\end{equation}
Simple MCMC calculations of the variance on Eq.\,\eqref{eq:effective_IP_estimator} confirm that this approximation works very well, with differences between values determined through Eq.~\eqref{eq:effective_IP_variance} and the MCMC variance differing by 5\% at most, always in lowest flux pixels. We later describe how we used the data themselves to modify $\sigma^2_{\hat{\psi}}$ where necessary.

\subsection{Obtaining line position and brightness}\label{sec:fitting_IP}
Accurate determination of $x_*$ and $f_*$ that go into calculating $\hat{\psi}$ requires that $\psi$ is already known and an accurate estimation of $\psi$, in turn, requires $x_*$ and $f_*$ to also be known. This motivated the use of an iterative approach.

Initial guesses for $x_*$ and $f_*$ were determined by approximating $\psi$ with a Gaussian profile and fitting it to the data with three free parameters: the amplitude $A$, the mean $\mu$, and the standard deviation $\tau$ of the Gaussian, that is
\begin{equation}\label{eq:gaussian_ip_fit_expression}
    I(x; A, \mu, \tau) = \frac{A}{ \tau\sqrt{2\pi}} \exp{\left[ -\frac{1}{2} \left(\frac{ x-\mu } {\tau}\right)^2\right]}. 
\end{equation}
Since we are interested in the effective IP, integration under pixel area when optimising for the Gaussian parameters was not performed. Instead, we simply evaluated the Gaussian profile at pixel centres. The impact this has on estimating $x_*$ is minimal (the average change in line centres is approximately \pix{1e-5}). We identify $x_*$ and $f_*$ with the mean and the integrated area under the Gaussian profile, that is $x_* = \mu$ and $f_* = A \tau\sqrt{2\pi}$.

In subsequent iterations, once $\psi$ has been modelled using our GP regression methods, the Gaussian approximation is replaced by the formula:
\begin{equation}\label{eq:empirical_ip_fit_expression}
    I(x;A_*, x_*, \omega) = A_* \psi \left(\omega(x-x_*)\right) .
\end{equation}
Here, $\psi $ is the IP normalised to unit area, and $A_*,x_*$, and $\omega$ are free parameters. $\psi(\Delta x)$ models were saved into a look-up table containing 800 values spanning $\Delta x =\pm\pix{8}$, corresponding to 50 values per pixel, and were interpolated using cubic splines during fitting. $A_*$ and $x_*$ are the line amplitude and central position, respectively, whereas $\omega$ is a parameter whose value is $\approx 1$, allowing for small stretching or compressing of the pixel scale when performing the fit. Because $\psi$ already describes integrated quantities, line brightness was calculated as $f_* = \sum_{i=1}^{N} I_i$, where the sum goes over the $N$ pixels of the line. 

\subsubsection{Fitting an IP model to the data}\label{sec:fitting_ip_to_data}

Let $I(\boldsymbol{\zeta})$ be the predicted model flux, where $\boldsymbol{\zeta}$ is the set of all free model parameters. Parameter optimisation finds $\boldsymbol{\zeta}$ which minimises the normalised residuals of the model to the data:
\begin{equation}\label{eq:least_squares_x}
    \boldsymbol{\zeta} = {\rm arg\,min} f(\boldsymbol{\zeta}), 
\end{equation}
where
\begin{equation}\label{eq:least_squares_f}
    f(\boldsymbol{\zeta}) = \sum_{j=0}^{j=N} \left( \frac{I(\boldsymbol{\zeta})_j - F_j}{\sigma_j} w_j \right)^2 .
\end{equation}
The summation in Eq.~\eqref{eq:least_squares_f} goes over the $N$ pixels of the astrocomb line, where the $j=0$ and $j=N$ correspond to the locations of the minima either side of the line and $w$ is the weight applied to each pixel. 

The weighting scheme was implemented to ensure the best possible fit at astrocomb line centres, following the discussion and recommendations made in \ak. The weight of a pixel centred at $x=j$ and at a distance $\Delta x = j - x_*$ from the astrocomb line, was given by
\begin{equation}\label{eq:weights}
{w_j = }
    \begin{cases} 
            {1} & \text{for } {|\Delta x| \leq x_1}\\
            {\frac{x_2-|\Delta x|}{x_2-x_1}} & \text{for } {x_1 < |\Delta x| \leq x_2}\\
            {0} & \text{for } {|\Delta x| > x_2},
    \end{cases}
\end{equation}
where $x_1=\pix{2.5}$ and $x_2=\pix{5}$. The weighting scheme was implemented when fitting both our empirical IP shapes and the Gaussian IP shapes.

The reported best-fit parameters correspond to the values of $\boldsymbol{\zeta}$ that satisfy Eq.~\eqref{eq:least_squares_x}. Parameter errors were calculated from the diagonal of the covariance matrix at the best-fit solution. 
\subsection{A GP for the instrumental profile}\label{sec:GP_for_IP}
The empirical $\psi$ was modelled using GP regression, conditioning the GP on the observed $\hat{\psi}(\Delta x)$ samples. Because the shapes of astrocomb lines are approximately Gaussian, the mean function was chosen to be
\begin{equation}\label{eq:gp_mean_function}
    m(x;\theta) =  A \; \mathcal{N}(x; \mu, \tau) + y_0
\end{equation}
where $\mathcal{N}$ represents the standard normal probability density function. This function takes parameters $\theta=( A, \mu, \tau, y_0)$ corresponding to the normalisation, the mean, the standard deviation, and the zero-offset, in that order.

The choice of covariance function affects the smoothness of the regression curve. The squared exponential covariance function was chosen, i.e.
\begin{equation}\label{eq:gp_correlation_matrix}
    k_\mathrm{SE}(x_i, x_j; \phi) = a^2 \exp \left( 
        -\frac{|x_i-x_j|^2}{2l^2}
    \right) ,
\end{equation}
which depends on the hyperparameters $\phi=(a,l)$ i.e. an amplitude and a length-scale. The amplitude controls how far the regression curve can stray from the mean function, while the length-scale determines typical length of wiggles in the curve. 

An additional diagonal term was added to the covariance matrix in order to account for the uncertainties on the data:
\begin{equation}\label{eq:gp_correlation_matrix+diagonal}
    k(x_i, x_j;\phi) = k_\mathrm{SE}(x_i, x_j; \phi) + \sigma_i\sigma_j \delta_{ij},
\end{equation}
where $\sigma_i = \sqrt{\sigma_{\hat{\psi}_i}^2 + \sigma_0^2}$, and $\sigma_0^2$ is an additional error term that is common to all $\hat{\psi}$ samples and is also a free hyperparameter. Finally, $\delta_{ij}$ is Kronecker delta.

\subsection{Defining the IP centre}\label{sec:line_centering}
Defining the location of the IP centre, that is deciding where to set $\Delta x=0$, is crucial for line centre $x_*$ measurements and, by extension, the construction of the IP models themselves. For an unimodal and symmetric IP profile, any of the mean, mode, or median can be meaningfully used as the centre. While these quantities are also well defined for asymmetric profiles, they can be very different from each other and also be sensitive to small changes in the shape of the profile.

We wanted our centre estimator be robust against small changes in hyperparameter values. We thus set to identify the optimal centre estimator for use in this work by comparing the properties the mean, the mode, the median, and the centre definition from \citet{Anderson2000}. In this last case, the IP centre is posited to be at the location which results in equal fluxes for the two brightest pixels: $\psi(\Delta x = 0.5) = \psi(\Delta x = -0.5)$. 

To find the most robust estimator, a single IP model was perturbed 1000 times and the four centre estimators were calculated for each perturbed model. The details of the comparison can be found in Appendix \ref{app:centre_estimator}. The results show that \ak's centre estimate is the most robust of the four examined, shifting by $\leqslant\pix{5e-8}$ across the 1000 perturbed profiles. For comparison, the mode was the least stable estimator with shifts $\geqslant\pix{1e-3}$.

We also followed \ak's suggestion to apply a rigid shift to the $\hat{\psi}$ samples in $\Delta x$ to ensure that the IP models are properly centred before modelling. The transformation applied was:
\begin{equation}\label{eq:anderson_shift}
    \Delta x \rightarrow \Delta x + \frac{\psi(0.5) - \psi(-0.5)}{\frac{\partial \psi}{\partial x} (-0.5) + \frac{\partial \psi}{\partial x} (0.5)}.
\end{equation} 
Subsequently to the data shifting, GP hyperparameters were optimised once more and the GP was retrained on the shifted data. The data shifting and hyperparameter optimisation procedures were repeated until the difference between consecutive shifts was smaller than \pix{1e-3} or for a maximum of twenty times (the stopping criterion was typically satisfied after three to four repetitions). This $\psi$ at the end of this procedure was the final model of a single iteration.

%
\section{Reconstructing the HARPS IP}\label{sec:IP_reconstruction}
%

\subsection{Data}\label{sec:instrument_data}

The methods described in the previous Sections were applied the High Accuracy Radial-velocity Planet Searcher (HARPS) spectrograph, and its IP was reconstructed from a single astrocomb calibration frame taken as a part of an observing programme to measure the fine structure constant ($\alpha$) at $z=1.15$ towards the quasar HE0515$-$4414 executed in December 2018 (ESO observing programme 0102.A-0697, PI Milakovi{\'c}). The primary reason for choosing HARPS is that the IP models could be applied to the science spectra in order to explore their impact on measurements of $\alpha$. The HARPS IP was previously investigated by \citet{Zhao2021}, although no IP shape models were produced. 

While ESPRESSO observations of the same quasar are also available and astrocomb calibrated, those observations required significant manual interventions in the data \citep{Murphy2022} for reasons not yet fully understood (ESPRESSO consortium, private communication). Additionally, the wavelength calibration of ESPRESSO contains systematic effects which are also not yet understood. The most obvious example is the high frequency correlation of its wavelength calibration residuals with wavelength. The residuals also are not normally distributed nor do they have properties expected from photon noise statistics \citep{Schmidt2021}. These problems have not been observed in HARPS.

\subsubsection{HARPS}
HARPS is a fibre-fed echelle spectrograph designed for extremely precise spectroscopic measurements \citep{Mayor2003}, installed at the ESO 3.6m telescope at the La Silla Observatory. HARPS's primary scattering element is an R4 grism which disperses the incoming light into 72 echelle orders, 89 through 161. 
Optical fibres transport the light from the telescope to the instrument in two channels. One channel (the primary, fibre A) generally carries the astronomical target light. The other channel (the secondary, fibre B) can be used to simultaneously record the sky light or the light from a wavelength calibration lamp. HARPS is equipped with three calibration lamps: a Thorium-Argon (ThAr) arc lamp, a Fabry-P{\'e}rot etalon, and an astrocomb, which can all be observed in either fibre. The nominal spectral resolution is $R=\SI{115 000}{}$. The entire instrument is enclosed in a temperature and pressure controlled vacuum vessel in order to minimise environmental disturbances and their impact on the scientific observations.

\begin{figure}
    \centering
    \includegraphics[width=\columnwidth]{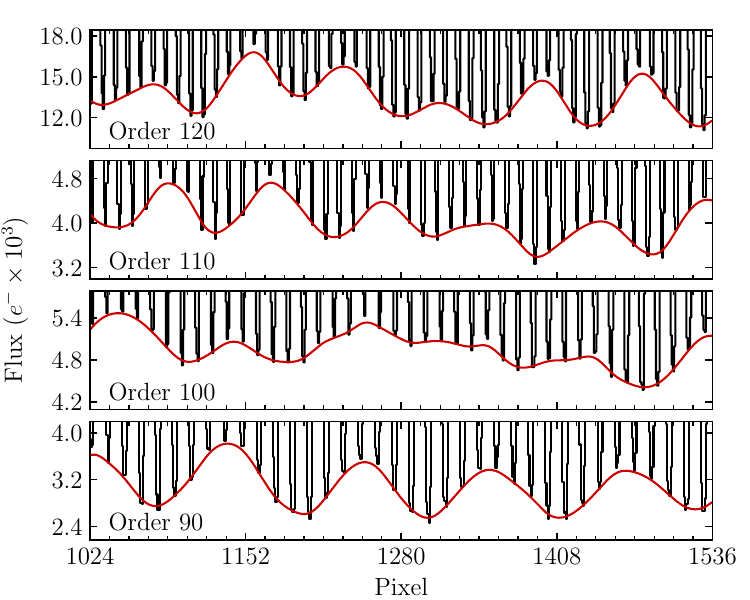}
    \caption{Astrocomb spectrum (black) and the inferred spectral background (red) for \pix{512} wide sections of four echelle orders.}
    \label{fig:background}
\end{figure}

\subsubsection{Astrocomb data description}
We first set out to use the astrocomb spectrum recorded on 2018-12-05 at 08:12:52.040 UT. That spectrum was used for wavelength calibration of quasar observations in \cite{Milakovic2021} and was thus the obvious choice. Upon closer inspection of that data, we discovered that the peaks of some astrocomb lines were saturated in that exposure, making it unusable for IP reconstruction. Instead, we used the spectrum recorded on 2018-12-07 at 00:12:50.196 UT, that had the second largest levels of flux in the set of calibration frames of this observing programme but the lines were not saturated. The offset and repetition frequencies of the astrocomb were $f_0=\SI{4.58}{\giga\hertz}$ and $f_r=\SI{18}{\giga\hertz}$. The reduced astrocomb spectrum (HARPS pipeline version 3.8) was downloaded from the ESO archive. The pipeline extracts the 1d spectra of each echelle order using optimal extraction \citep{Horne1986,Robertson1986}, producing separate files for the two channels. We corrected the flux for the detector gain and for the measured spectrograph blaze function\footnote{The relevant blaze function is saved in the observations taken on 2018-12-06 at 21:50:22.394 UT}. 

The spectral variance array was not saved by the pipeline so we estimated it from the flux array assuming Poissonian statistics, i.e. $\sigma_{F_i}^2 = F_{i}$, where $F_{i}$ is the flux in $i^{\rm th}$ pixel in electrons. Given the high flux values in the astrocomb spectra, the contributions from the dark current and the read-out process were assumed to be negligible.

\subsection{Background estimation}\label{sec:background}
The spectral background contains contributions from scattered light \citep[$\approx 3\%$, ][]{HARPS_manual} and from background associated with the astrocomb system itself. The astrocomb background is seeded by the high-power amplifier of the system required to broaden the astrocomb spectrum from the infrared to the optical wavelengths inside the photonic crystal fibre \citep{Probst2015,Probst2020}, and contributes as much as 30\% of the total flux below \SI{500}{\nano\meter} \citep{Milakovic2020}. 

The full spectral background was determined in each echelle order independently. We started by identifying the minima as the locations at which the first derivative of the spectral flux array is closest to zero and, simultaneously, second derivative of the flux array is positive. Before calculating the derivatives, the flux array was upsampled by a factor of 10 and smoothed using a Nuttall window function \citep{Nuttall1981} to more precisely determine the location of the minima and to avoid falsely detecting minima in the noise. The most appropriate width for the window function was determined by calculating the power spectrum of the flux array and identifying the frequency carrying the most power. In practice, the window widths were between 11 and 15 pixels, depending on the wavelength. The background over the entire order was then determined by connecting the minima between the lines using straight lines and smoothing the resulting array with a Nuttall window function with a window size of \pix{51}. Small sections of the background is shown in \figref{fig:background} for four echelle orders. We assumed that the background follows Poisson statistics, that is $\sigma^2_{B_i} = B_i$.

\subsection{Line detection and line properties}\label{sec:line_detection}
Boundaries between individual astrocomb lines were identified by applying the minima detection algorithm to the background-subtracted spectra. 10576 astrocomb lines were detected, with mode numbers between 24087 and 33384 (1472 modes falling in the spectral region covered by two adjacent orders are seen twice). The separation between lines ranges between \pix{10} at \SI{500}{\nano\meter} and \pix{18} at \SI{690}{\nano\meter}, whereas the FWHM of the best fitting Gaussian for each line was <\pix{3.7} everywhere. Lines were therefore sufficiently separated such that line blending is expected to be negligible. This provided $> \SI{100000}{}$ samples of $\hat{\psi}(\Delta x)$ across a 60\% of the detector area.

Visual examination of lines revealed asymmetry in their shapes, with most lines being skewed to the right, and this asymmetry also appeared to vary across the detector. We confirmed this by examining differences between two line centre estimates: the centroid ($\overline{x}$, a flux-weighted average position of the line) and the centre of the best fitting Gaussian, $x_{*,{\rm G}}$. If the lines are approximately symmetric, there should be zero difference between these two centre estimates whereas positive or negative values imply that the line is skewed to the right or to the left, respectively.

\begin{figure}
    \centering
    \includegraphics[width=\columnwidth]{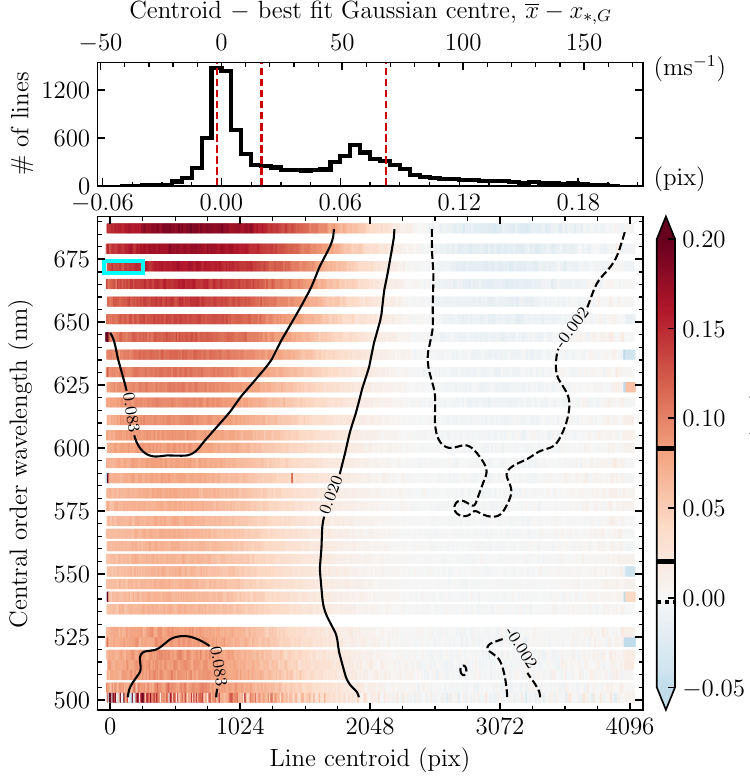}
    \caption{Astrocomb line shape asymmetry as a function of position on the HARPS detector. Main panel: Each coloured square shows one of the 10576 detected astrocomb lines in the HARPS astrocomb spectrum described in the text. The position of the square in the $(x,y)$ plane approximately matches its position on the HARPS detector frame. The gap seen at $\approx \SI{525}{\nano\meter}$ is due one echelle order falling between the two HARPS detector CCD chips so is not recorded. Wavelength within a single echelle order increases towards the right. The colour of the square is proportional to the difference (in units pixels) between the measured line centroid and the line position determined by fitting a Gaussian IP to the data. The zero point of the colour bar to the right is set to zero, such that the blue and the red colours denote negative and positive differences, respectively. Smooth variations across the detector most likely correspond to IP shape variations. Contours over smoothed data are plotted to visualise trend more clearly and emphasise where certain values cluster. Contour levels correspond to the 16\ts{th}, 50\ts{th}, and 68\ts{th} percentile of the values, i.e., \pix{-0.002}, \pix{0.020}, and \pix{0.083}. Full lines are used for positive levels and dashed lines for negative. The cyan rectangle indicates the location of the $1^{\rm st}$ segment in optical order 90, for which we show an empirical IP model in \figref{fig:IP_example}. 
    Top panel: Histogram of the values shown in the main panel. The secondary $x$-axis on the top shows the line centre differences in units \mps{} ($\pix{1}=\mps{820}$). The vertical dashed red lines show the positions of the distribution median and the central 68\% distribution limits. The same quantities are shown as thick black lines in the colour bar.
    }
    \label{fig:CCD_centroid-gaussian}
\end{figure}
The main panel of \figref{fig:CCD_centroid-gaussian} shows values of this quantity for all detected lines as coloured squares. Lines appearing in a single order are aligned along the $x$-axis and echelle orders are separated along the $y$-axis, so that the location of the squares in the panel approximately corresponds to their location on the detector. Wavelength within each order increases with increasing pixel number (left to right), whereas the central wavelength of the order increases (order number decreases) from bottom to top. 

Correlated departures from a symmetrical shape across the detector are immediately noticeable. The most obvious is the division of the detector area into two halves: left, with positive $\overline{x}-x_{*,{\rm G}}$ values (coloured red) and right, with values around zero (white) or negative (light blue). Lines appearing in the middle of each order have values closest to zero. Maximal positive departures from the symmetrical shape are seen in the top left corner of the panel, with $\overline{x}-x_{*,{\rm G}}$ as large as \pix{0.14}, and maximal negative departures are seen in the top right corner of the panel, with values of approximately \pix{-0.03}. In between the two regions the values vary quickly. Negative values seem to appear around \pix{3000} in most orders, but in some orders $\overline{x}-x_{*,{\rm G}}$ recovers to zero at the right edge. The width of regions with negative values seem to increase as one moves vertically from the middle of the detector towards the top and the bottom. Contours in the same Figure illustrate this more clearly. We conclude from this that the HARPS IP is not symmetric and that its shape varies strongly across the detector in a correlated way.

The top panel of \figref{fig:CCD_centroid-gaussian} shows the histogram of the values from the main panel. The mode of the distribution is around zero pixels, and its median is \pix{0.020}. The central 68\% of the distribution lies within \pix{-0.002} and \pix{0.083}. A single HARPS pixel covers a wavelength range that corresponds to a Doppler velocity shift of approximately \mps{820}, so the same quantities can be expressed as velocity shifts. Converting the median and the central 68\% distribution limits into velocities, we obtained \mps{16.3}, \mps{-1.7}, and \mps{67.7} (respectively). Out of the 10576 lines, 4266 of them (40\%) have values within $\pm \pix{0.01}$ (\mps{8.2}).

\subsection{Determining the IP locally}\label{sec:division_into_segments}
To reduce the impact from positional IP variability on our reconstructed models as much as possible, the detector was artificially segmented in both the main and the cross dispersion directions and the IP models was then determined for each segment independently. Echelle orders provide a natural division in the cross dispersion direction and we further divided individual orders into 16 segments, each spanning \pix{256} or $\approx \SI{0.5}{\nano\meter}$, in the main dispersion direction. The 256 pixel width was chosen to avoid a known ``stitching'' issue of the HARPS detector which appears as small changes in the detector pixel size every \pix{512} in the main dispersion direction and arises from imperfections during CCD construction \citep{Wilken2010,Bauer2015,Coffinet2019,Milakovic2020}. The area covered by one such segment is shown as a cyan rectangle in the top left corner in the main panel of \figref{fig:CCD_centroid-gaussian}.

\begin{figure}
    \centering
    \includegraphics[width=\columnwidth]{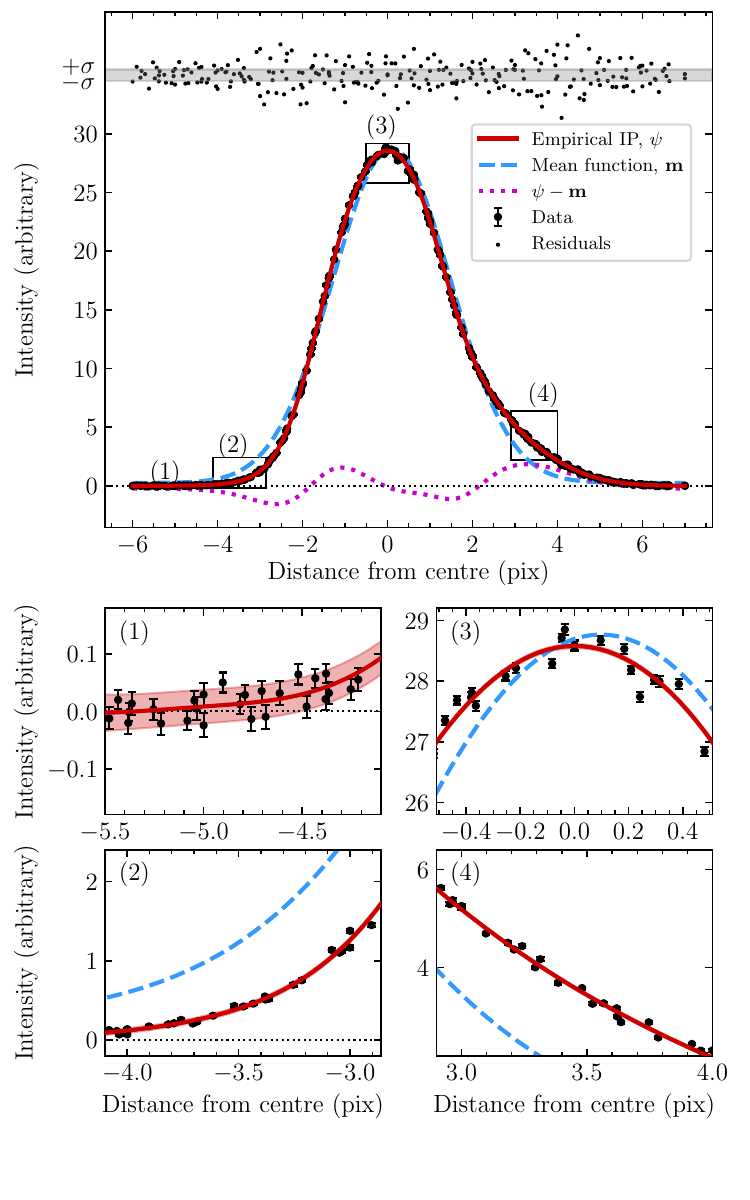}
    \caption{GP fits to the IP. Top panel: The large black points are the 251 $\hat{\psi}$ samples with corresponding error bars in the first segment of order 90 (average wavelength $\lambda=\SI{676.5}{\nano\meter}$), also indicated as a rectangle in \figref{fig:CCD_centroid-gaussian}. The thick red line is our reconstruction of $\psi$ using GP regression (\secref{sec:GP_regression}) after re-centring, in the first iteration. The secondary GP for variance estimation was not used, resulting in a $\chi^2_\nu=5.01$ for this fit. The shaded red bands show the $1\sigma$ ranges of the model. The dashed blue line is the mean function of the GP, $\mathbf{m}(x;\theta)$, Eq.~\eqref{eq:gp_mean_function}. The small black points above the model show the residuals normalised with respect to the error on the data, with the grey shaded region indicating $\pm1\sigma$. Parameters $\theta$ (controlling $\mathbf{m}$) and $\phi$ (controlling the correlation matrix $\mathbf{K}_{ij}$, Eq.~\eqref{eq:gp_correlation_matrix}) were fitted simultaneously. The magenta dotted line is the difference between the GP model and the mean function, that is the function controlled by $\mathbf{K}_{ij}$. Departures from the Gaussian shape are observed almost everywhere. The thin dotted black line indicates zero flux. Rectangles indicate regions plotted in the bottom panels. Panels with numbers (1)-(4): Zoom-ins of the top panel. The number in the top left corner identifies one rectangle in the top panel. }
    \label{fig:IP_example}
\end{figure}

While we explicitly assumed that IP does not change within the segment when performing the modelling, we allowed for smooth variation of $\psi$ when fitting IP models to the data, as follows. Let the sets $\{\psi_1, \ldots , \psi_{16}\}$ and $\{X_1, \ldots, X_{16}\}$ denote the 16 IP models calculated within single echelle order and the central segment positions, respectively. We accepted the models as correct at those locations, $\psi|_{x=X_i} \equiv \psi_i$, where $i$ is the ordinal number of the segment and $x$ is now a coordinate within the echelle order and not $\Delta x$. The IP at an arbitrary $x$ was then interpolated from the two closest $\psi$ models by applying weights inversely proportional to the distance between $x$ and the two nearest segment centres:
\begin{equation}\label{eq:IP_mixing}
    \psi (\Delta x)\big\rvert_{X_i<x<X_{i+1}} = \frac{d_2}{D} \psi(\Delta x)\big\rvert_{X_i}  + \frac{d_1}{D} \psi(\Delta x)\big\rvert_{X_{i+1}}.
\end{equation}
Here, $d_1 = |x - X_i|$,  $d_2 = |x - X_{i+1}|$, and $D=|X_{i+1}-X_i|$. If $x<X_1$ or $x>X_{16}$, no interpolation was done and only the single closest IP was used ($\psi_1$ or $\psi_{16}$).

\subsection{Hyperparameter determination}\label{sec:hyperparameter_determination}
To model the IP for one segment, we optimise $\mathcal{L}$ from Eq.\,\eqref{eq:likelihood} using $\hat{\psi}$ samples falling within the segment boundaries to find the maximum likelihood values of the hyperparameters $\theta$ and $\phi$. The mathematical framework for hyperparameter optimisation was implemented using the GP library \texttt{tinygp} \citep{tinygp}, which was performed using the limited memory Broyden - Fletcher - Goldfarb - Shanno algorithm with boxed constraints \citep[also known as L-BFGS-B,][]{Byrd1995} as implemented in the \texttt{Python} optimisation library \texttt{jaxopt} \citep{jaxopt}. After this step, known as training, we can draw sample functions $f$ from the GP evaluated at arbitrary locations, and the variation of these samples give us an estimate of the uncertainty in the regression function $f$. The mean $f$ is the IP model, $\psi$.

During training, parameters $( A, \mu, \tau, y_0)$ were initialised at the values determined from fitting Eq.~\eqref{eq:gp_mean_function} to $\hat{\psi}(\Delta x)$ using non-linear least-squares. These parameters were constrained not to move further than $\pm5$ standard deviations, which were calculated from the covariance matrix diagonal at the best fit solution. We required hyperparameters $a$, $l$, and $\sigma_0$ to always be positive so they were expressed as logarithms. Their initial values were $\log a = 1$, $\log l = 0$, and $\log \sigma_0 = -5$. These parameters were constrained to have the following values: $-4 \leqslant \log a\leqslant 4 $, $-1 \leqslant \log l \leqslant 2 $, and $-15 \leqslant \log \sigma_0 \leqslant 1.5$. Only $\sigma_0$ ever reaches its imposed boundary (on the lower side). 

We also explored a fully Bayesian treatment, whereby we introduce hyperpriors on $\theta$ and $\phi$ and perform Markov Chain Monte Carlo (MCMC) sampling of the posterior distribution, thus propagating additional uncertainty to the regression function. We found MCMC sampling prohibitively slow for general use (several hours per segment using MCMC compared to one minute per segment using L-BFGS-B). However, our tests suggest that our maximum likelihood uncertainty estimates are very similar to the fully Bayesian approach and the IP models derived through L-BFGS-B are indistinguishable from the MCMC one. More details are provided in Appendix \ref{app:mcmc}.

\subsection{Empirical variance estimation}\label{sec:empirical_variance}

We already commented that our $\sigma_{\hat{\psi}}$ estimates from Eq~\eqref{eq:effective_IP_variance} are formally incorrect. \figref{fig:IP_example} shows the model derived from $\hat{\psi}$ samples in the $1^{\rm st}$ segment of order 90 (wavelength range $\SI{676.1}{\nano\meter}\leqslant\lambda\leqslant\SI{676.7}{\nano\meter}$) in detail. While the fit is generally good, the residuals scatter beyond what is expected from errors on individual data points. This is most obviously seen in the peak region, where the residuals show correlated scatter outside of the expected $\pm1\sigma$ range indicated by the grey shaded area in the top of the Figure's main panel. A detailed view, shown in panels (1)-(4) of the same Figure, reveals that the scatter between nearby $\hat{\psi}$ samples is inconsistent with the scatter expected from their estimated errors and that this inconsistency changes with $\Delta x$. For example, panel (3) shows that the errors in the region close to the profile peak are underestimated. 

The increased scatter also affects the goodness of fit statistic, reduced $\chi^2$: $\chi^2_\nu = \chi^2/\nu$, where $\nu$ is the number of degrees of freedom in the fit. $\chi^2$ is the weighted sum of squared deviations from the expected value
\begin{equation}\label{eq:chisq}
    \chi^2 =  \sum_{i=1}^N \left( \frac{  O_i - E_i}{\varsigma_i } \right)^2,
\end{equation}
where $O_i$ and $E_i$ are the observed and the expected values of the $i^{\rm th}$ observation, and $\varsigma_i$ is its standard error.
In this case, $O_i=\hat{\psi_i}$, $E_i=\psi(\Delta x_i)$, $\varsigma_i = \sigma_i$, and the summation goes over the $N$ samples of a single segment. For the model shown in \figref{fig:IP_example}, $\chi^2_\nu=5.01$, significantly higher than the desirable value of $\approx 1$. Models for other segments had similar $\chi^2_\nu$ values, most likely due to incorrect errors on $\hat{\psi}$. Even though we also obtained models with $\chi^2_\nu\approx1$, strong departures of the normalised residuals from the $\pm1\sigma$ range, such as those seen in \figref{fig:IP_example}, were observed also in those models.

\begin{figure}
    \centering
    \includegraphics[width=\columnwidth]{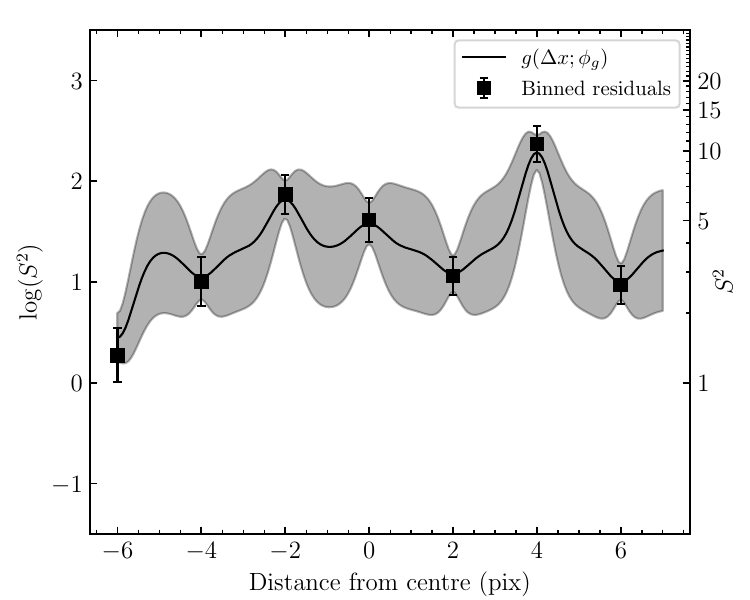}
    \caption{GP for empirical variance estimation. $g(\Delta x)$ multiplies $\sigma_{\hat{\psi}}$ to capture the observational variances more correctly. Black squares with error bars are the logarithms of $S^2$, where $S^2$ is calculated as explained in the text. The solid black line is the mean prediction of $g(\Delta x)$ from GP regression and the grey shaded are the $1\sigma$ ranges. The $y$-axis label on the right shows $S^2$ and $g(\Delta x)$ on a linear scale.}
    \label{fig:scatter_gp}
\end{figure}

\subsubsection{A secondary GP for variances}
Since these models underestimated the observational variance, the resulting uncertainty estimates on line positions were also likely to be underestimated. We therefore implemented additional methods to deal with this issue by extending our modelling procedures to more correctly capture the observed variance. A second GP was introduced to modify $\sigma_i$ from Eq.~\eqref{eq:gp_correlation_matrix+diagonal}, of the form
\begin{equation}\label{eq:gp_for_scatter}
    g(\Delta x) \sim \mathcal{N}(C_g,\mathbf{G}_{i,j}), 
\end{equation}
where $C_g$ is a constant hyperparameter and $\mathbf{G}$ is a squared exponential kernel with the amplitude $a_g$ and the length-scale $l_g$ as hyperparameters. The mean prediction of $g(\Delta x)$ acts as a multiplicative factor modifying the observed variance of $\hat{\psi}_i$ such that $\sigma_i$ going into Eq.~\eqref{eq:gp_correlation_matrix+diagonal} becomes:
\begin{equation}\label{eq:epsilon_i_modified}
    \sigma_i = \sqrt{g_i\sigma_{\hat{\psi}_i}^2 + \sigma_0^2},
\end{equation}
where $\sigma_{\hat{\psi}_i}^2$ is the unmodified error on $\hat{\psi}_i$ and $g_i= g(\Delta x_i)$.

The training data for $g$ was derived from the normalised model residuals, $(\hat{\psi}-\psi)/\sigma_{\hat{\psi}}$, as follows. In the first step, the $N$ residuals were divided into equally spaced bins in $\Delta x$ such that there were at least 15 points in every bin. Sample variance, $S^2$, was then calculated together with the variance on $S^2$ for each bin:
\begin{equation}\label{eq:sample_variance_variance}
    {\rm Var}(S^2) = \frac{1}{n} \left (\mu_4 - \frac{n-3}{n-1} S^4 \right).
\end{equation}
Here, $n$ is the number of residuals in the bin and $\mu_4$ is their fourth central moment. 

Hyperparameters $\phi_g=\{C_g, a_g,l_g\}$ were then optimised using equations in \secref{sec:GP_regression} on pairs of points $(x_b,\log(S^2))$, where $x_b$ are bin centres. Limits on $\phi_g$ were: $-10 \leqslant \log C_g \leqslant 5 $, $-3 \leqslant \log a_g \leqslant 3 $, and $-1 \leqslant \log l_g \leqslant 3$. Logarithms were used to ensure $g(x)$ and hyperparameters $\phi_g$ are always positive. The value $\log({\rm Var}(S^2))$ was added to the diagonal of $\mathbf{G}$ to account for the uncertainty on the variances analogously to Eq.~\eqref{eq:gp_correlation_matrix+diagonal}. An example of $g(\Delta x)$ determined in this way is shown in \figref{fig:scatter_gp}, using the data and the model previously shown in \figref{fig:IP_example}. This determination of $g(\Delta x)$ is also in excellent agreement with MCMC determinations (Appendix \ref{app:mcmc}).

Finally, hyperparameters of the GP describing the instrumental profile were recalculated using the empirical variances. Appendix \ref{app:ip_example} shows a detailed plot for the same data as in \figref{fig:IP_example} obtained in this way. The multiplication function applied to the variances on $\hat{\psi}$ is the same as shown in \figref{fig:scatter_gp}. The improvement from applying this method is obvious: the residuals have improved significantly and show no dependence on $\Delta x$ and the value of $\chi^2_\nu$ is much closer to unity. Similar improvements are seen for all segments. 

\subsection{The final IP models}

We produced ten iterations of IP models for each of the 16 segments of the 32 echelle orders illuminated by the astrocomb (5120 models in total). In each iteration, all astrocomb lines were refitted using the $\psi$ models from the preceding iteration, where the IP shape was also interpolated to the location of the astrocomb line, as explained in \secref{sec:division_into_segments}. Surprisingly, we found that the likelihood of the model does not necessarily increase with each subsequent iteration. Furthermore, models with lower likelihoods were also associated with worse fits to the data and larger uncertainties on model parameters. We thus also separately saved a set of 512 ``most likely'' $\psi$ models from the set of 5120 by choosing the model with the largest likelihood from the set of 10 models produced for one segment. This set of models also results in the average $\chi^2_\nu$ (calculated over all astrocomb lines) being the closest to unity.

\begin{figure*}
    \centering
    \includegraphics[height=0.93\textheight,keepaspectratio]{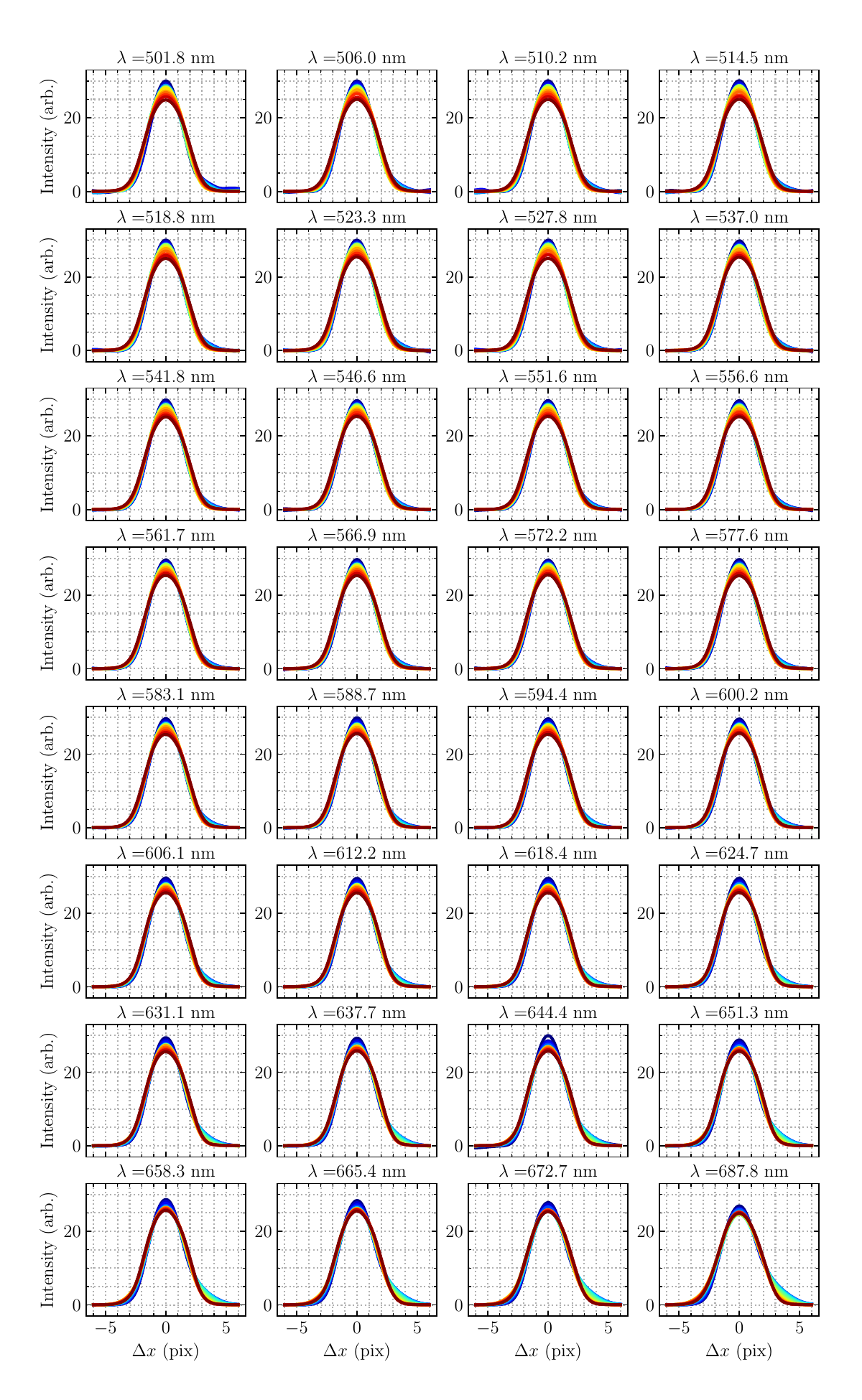}
    \caption{HARPS IP models in pixel space for orders 89 through 122. Each panel shows the IP models for a single echelle order, whose central wavelength is printed above it. The 16 coloured lines inside each panel show how the IP changes with wavelength (line colour changes from blue to red with increasing wavelength). }
    \label{fig:IP_all_pixel}
\end{figure*}

\begin{figure}
    \centering
    \includegraphics[width=\columnwidth]{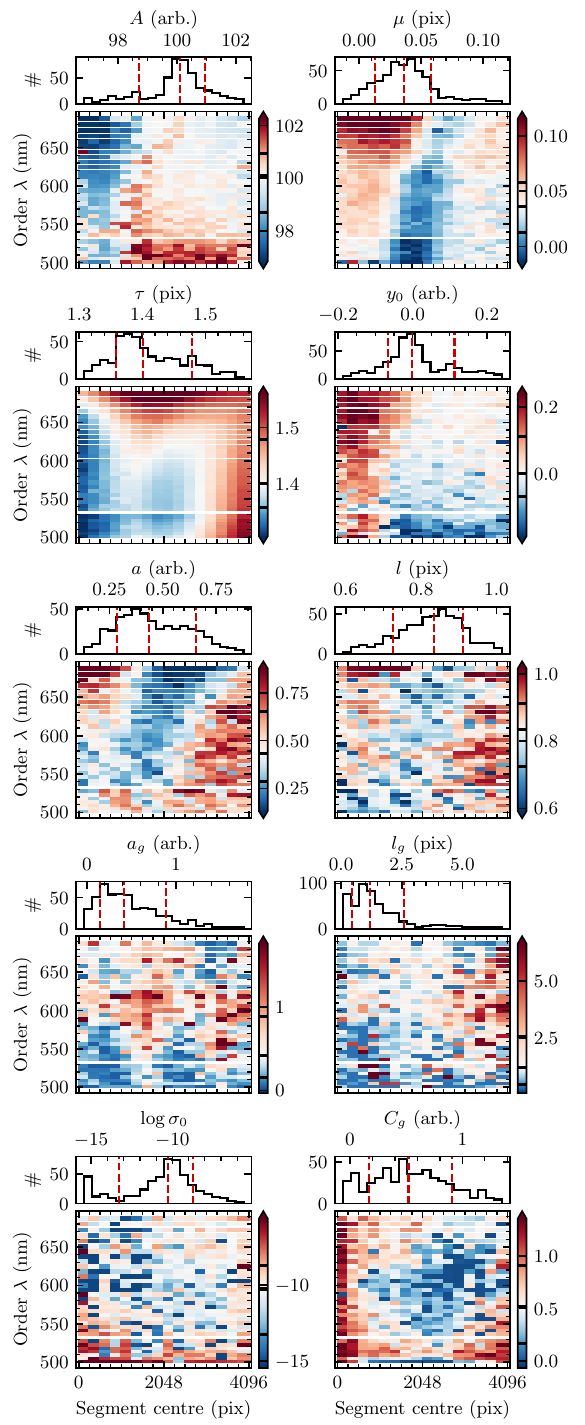}
    \caption{Values of hyperparameters describing $\psi$ as a function of detector position, also plotted as histograms above the main panels. White colour in the main panels corresponds to the median hyperparameter value. To better visualise trends, the colour bar and histogram limits only show the central 95\% of the plotted values. The units on the colour bar are indicated above the panels. The vertical dashed red lines showing the median and the central 68\% limits (calculated over all values) which are also shown as thick black lines in the colour bar. }
    \label{fig:GP_hyperparameters_mosaic}
\end{figure}

IP models in pixel space for all 32 orders are shown in \figref{fig:IP_all_pixel}, where one order is shown per panel. Looking at any of the panels, the strong variation of the IP shape within the order is immediately noticeable, with the bluest line (corresponding to the segment covering the bluest wavelength range) generally being the most sharply peaked and the reddest line (corresponding to the reddest wavelength covered by that order) being more broad. The IPs in the reddest echelle orders exhibit a tail on their right side that is the strongest for blue lines and that almost completely disappears for the reddest line, changing smoothly with wavelength. Inter-order variations are best appreciated by comparing the bluest lines in each panel, revealing that the aforementioned tail grows with increasing wavelength.

The most likely values for all ten hyperparameters are shown as a function of detector position in \figref{fig:GP_hyperparameters_mosaic}, together with their histograms (above each panel). Most notable is that most (if not all) hyperparameters change smoothly across the detector area, albeit the degree of smoothness varies between them. We also noted similarities between spatial distributions of some hyperparameters, such as between $\mu$ and $y_0$, and $a$ and $l$. The distribution of $\mu$ is consistent with what is seen in \figref{fig:CCD_centroid-gaussian}, where the lines in top left corner of the detector showed the strongest signs of asymmetry in their shapes and lines in the right detector edge were more symmetric, albeit not fully. Similar behaviour is seen in all echelle orders, once more consistent with the general picture on line shapes provided by \figref{fig:CCD_centroid-gaussian}.

\begin{figure*}
    \centering
    \includegraphics[width=\textwidth]{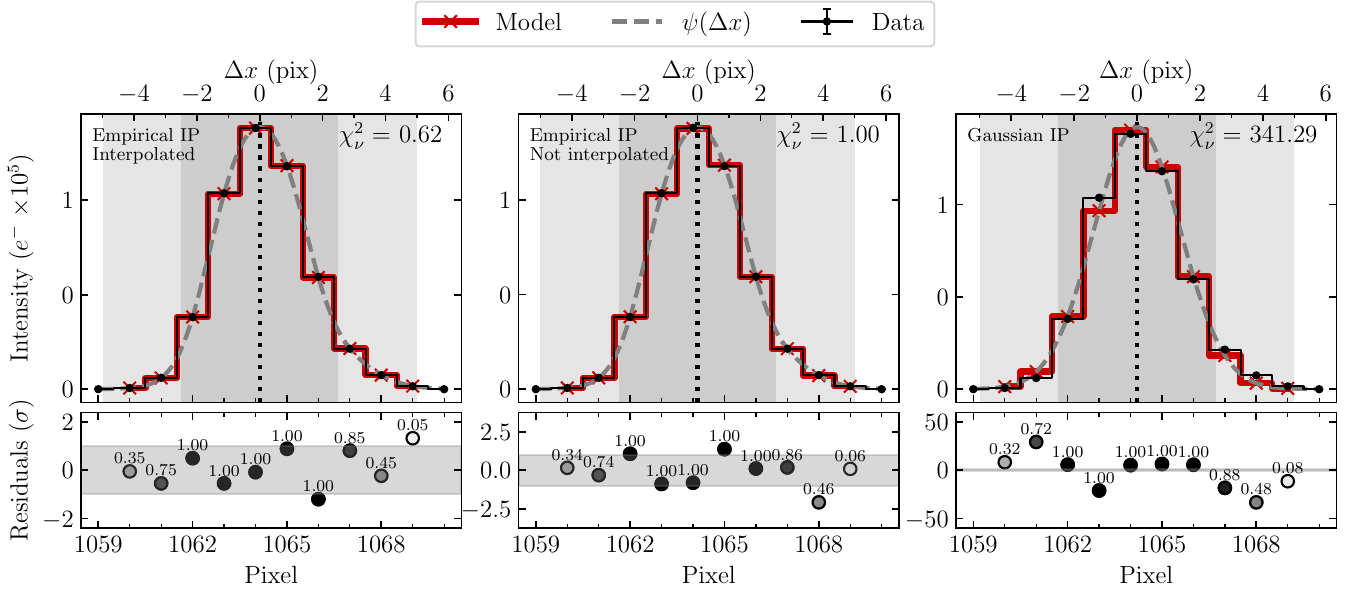}
    \caption{Results of fitting the same astrocomb line using three different IPs. They are, in order from left to right, (1) the most likely empirical IP, allowed to vary locally within the echelle order (interpolated), (2) when the empirical IP from the nearest segment is used instead (not interpolated), and (3) the Gaussian IP. In each top panel, the solid black histogram with error bars shows the astrocomb line flux and the solid red line with crosses shows the best fit model of that data. The line model is described by Eq.\,\eqref{eq:empirical_ip_fit_expression} when fitting the empirical IP and by Eq.\,\eqref{eq:gaussian_ip_fit_expression} when fitting a Gaussian IP. The location of the best-fit line centre, $x_*$, is indicated by the vertical dotted black line in each panel, with the ticks and labels on the top indicating $\Delta x$. The two vertical grey shaded bands illustrate the limits of the weighting scheme, Eq~\eqref{eq:weights}. $\chi^2_\nu$ for each fit is printed in the top right corner of the panel. In calculating $\chi^2_\nu$, no data weighting was applied. Finally, the continuous dashed grey line shows the IP model, $\psi$, shifted and rescaled to match the data (for visual comparison purposes only).
    Dots in each bottom panel are the normalised fit residuals and the grey shaded band indicates the $\pm1\sigma$ region. Dot opacity is proportional to the pixel weight in the fit (Eq.\,\eqref{eq:weights}), which is also printed above each dot. Dots falling within the darker grey band in the top panel all had a weight of unity, whereas the dots falling outside of both bands had zero weight (so are not plotted). Dots falling in between, that is inside the lighter grey band of the top panel, had weights that change linearly between zero and unity. We note that the $y$-axis scale differs between the bottom panels.}
    \label{fig:LSF_fit_single_line_good}
\end{figure*}
%
\section{Improving wavelength calibration accuracy}\label{sec:calibration_results}
%
\subsection{Fit quality of astrocomb lines and uncertainties on line position measurements}\label{sec:fit_quality}

\figref{fig:LSF_fit_single_line_good} shows the models for one randomly chosen astrocomb line fitted with three different IPs: (1) the empirical IP at the line location (i.e.\, $\psi$ was linearly interpolated from the nearest two segment centres), left panel; (2) the empirical IP from the nearest segment centre (no $\psi$ interpolation was done), middle panel; and (3) a Gaussian IP, right panel. The best fit quality, in terms of $\chi^2_\nu$, is provided by the interpolated empirical IP, followed by the empirical IP without interpolation, and finally by the Gaussian IP. The non-interpolated empirical IP is shown here for comparison and is not used elsewhere. We note the large difference in $\chi^2_\nu$ between the Gaussian IP model fit and either one of the two empirical IP model fits, demonstrating again that the Gaussian IP is inadequate. Similar results are obtained for all other examined lines. 

\begin{figure}
    \centering
    \includegraphics[width=\columnwidth]{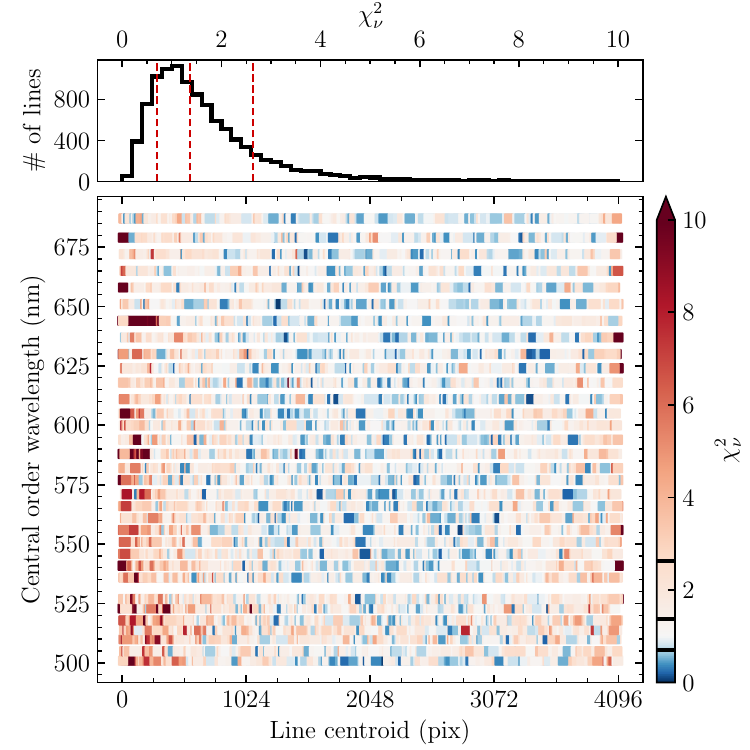}
    \caption{Reduced $\chi^2$ for 10576 fits of astrocomb line shapes using our empirical IP models. Main panel: $\chi^2_\nu$ as a function of position on the detector. The colour bar's zero point corresponds to $\chi^2_\nu = 1$, such that blue and red colours show lines whose $\chi^2_\nu$ is smaller and larger than 1 (respectively). Top panel: Histogram of $\chi^2_\nu$ values plotted in the main panel. The vertical dashed red lines show the locations of the median and the central 68\% distribution limits. The same quantities are shown as thick black horizontal lines in the colour bar.}
    \label{fig:LSF_chisqnu_ccd}
\end{figure}

All 10576 astrocomb lines were independently fitted using our empirical IP shapes and a Gaussian IP shape, yielding two sets of model parameters, including line centres (see Eq.\, \eqref{eq:gaussian_ip_fit_expression} and \eqref{eq:empirical_ip_fit_expression}). In the following text, we use subscripts to indicate which model parameters are referred to (``E'' for our empirical IP models and ``G'' for Gaussian IP models). The $\chi^2_\nu$ for all lines fitted using the final empirical IP models are shown in \figref{fig:LSF_chisqnu_ccd} as a function of line position on the detector. No data weighting was applied in calculating $\chi^2_\nu$. 

Looking at the spatial distribution of values (main panel), we see that the overall fit quality is excellent although lines closer to order edges have systematically higher $\chi^2_\nu$ values than lines in the middle of the detector. This indicates issues in those regions relating either to our models or to the observed fluxes and their uncertainties. Inaccuracies in the blaze correction function would adversely impact both on the construction of $\psi$ models (by increasing scatter between neighbouring $\hat{\psi}$ samples) and on the fitting procedure (by providing incorrect spectral fluxes and errors). Upon inspection, we found significant correlation between the blaze correction function and larger $\chi^2_\nu$ values.

As seen from the histogram in the Figure's top panel, the most commonly observed $\chi^2_\nu$ values is in the range $0.9\leqslant\chi^2_\nu\leqslant1.1$ and the median $\chi^2_\nu=1.41$. The central 68\% of the distribution is within $0.73\leqslant \chi^2_\nu \leqslant2.67$, with 70\% (95\%) of the values $\chi^2_\nu < 1.96\, (4.19)$. For comparison, the median $\chi^2_\nu$ for the Gaussian fits is 155.91, and 70\% (95\%) of the lines have $\chi^2_\nu < 186.22\, (270.30)$. Overall, the fit quality is significantly better when using our $\psi$ models than when using a Gaussian IP shape approximation. The improvement is also appreciated by comparing our \figref{fig:LSF_fit_single_line_good} and \ref{fig:LSF_chisqnu_ccd} to figures 4 and 5 in \citet{Milakovic2020}.

\begin{figure}
    \centering
    \includegraphics[width=\columnwidth]{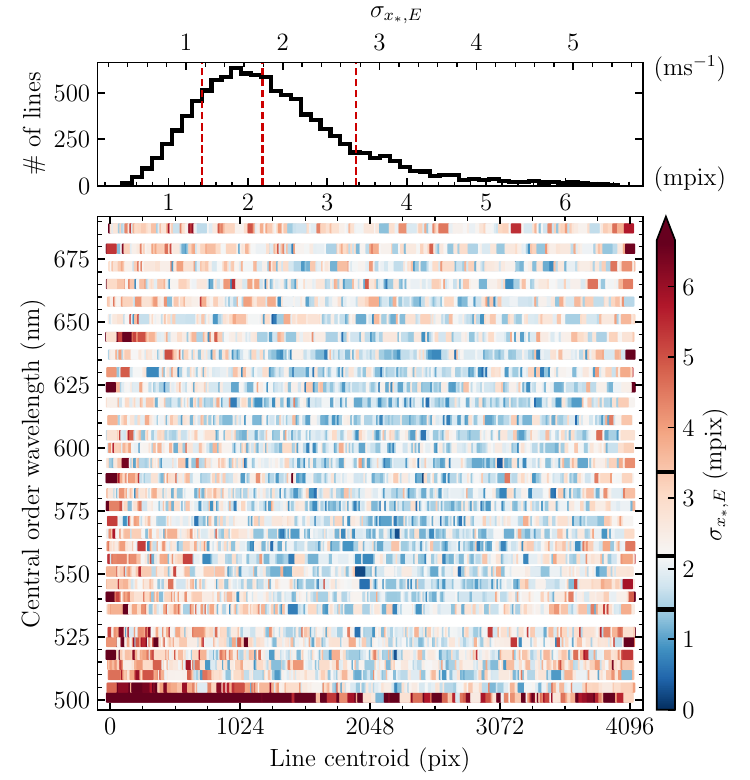}
    \caption{Uncertainty on astrocomb line centres from fitting our empirical IP models to the data, $\sigma_{x_*,E}$. Main panel: $\sigma_{x_*,E}$ as a function of position on the detector. The colour bar zero point is set to the sample median, such that the red and the blue colours correspond to values larger and smaller than the median, respectively. Lines with larger uncertainties generally have larger  $\chi^2_\nu$ (c.f.\ \figref{fig:LSF_chisqnu_ccd}) or low S/N (at \SI{500}{\nano\meter}). Top panel: Histogram of the values plotted in the main panel. The vertical dashed red lines show the median and the central 68\% distribution limits, which are also shown as thick black horizontal lines in the colour bar. The $x$-axis on the top shows $\sigma_{x_*}$ in units \mps{} ($\pix{1}=\mps{820}$). }
    \label{fig:CCD_lsf_xstar_uncertainty}
\end{figure}

\figref{fig:CCD_lsf_xstar_uncertainty} shows the uncertainties on line positions, $\sigma_{x_*,E}$, determined from the covariance matrix at the best-fit solution. As expected, larger uncertainties are associated with lines with higher $\chi^2_\nu$ values (e.g.\ at detector edges) and in regions with low S/N in the data. The median uncertainty is \SI{2.2}{\milli\pixel} and the central 68\% distribution limits are \SI{1.4}{\milli\pixel} and \SI{3.4}{\milli\pixel}. 95\% of all lines have $\sigma_{x_*,E}\leqslant \SI{4.9}{\milli\pixel}$. Approximating $\sigma_{x_*,E}$ in units \mps{} ($\pix{1}= \mps{820}$), distribution median corresponds to \mps{1.79} and the 16\ts{th}, 68\ts{th}, and 95\ts{th} percentiles correspond to \mps{1.17}, \mps{2.76}, and \mps{4.01}, respectively.

\subsection{Differences between Gaussian and empirical IP centres and the impact on wavelength calibration}\label{sec:wavelength_calibration}

\begin{figure}
    \centering
    \includegraphics[width=\columnwidth]{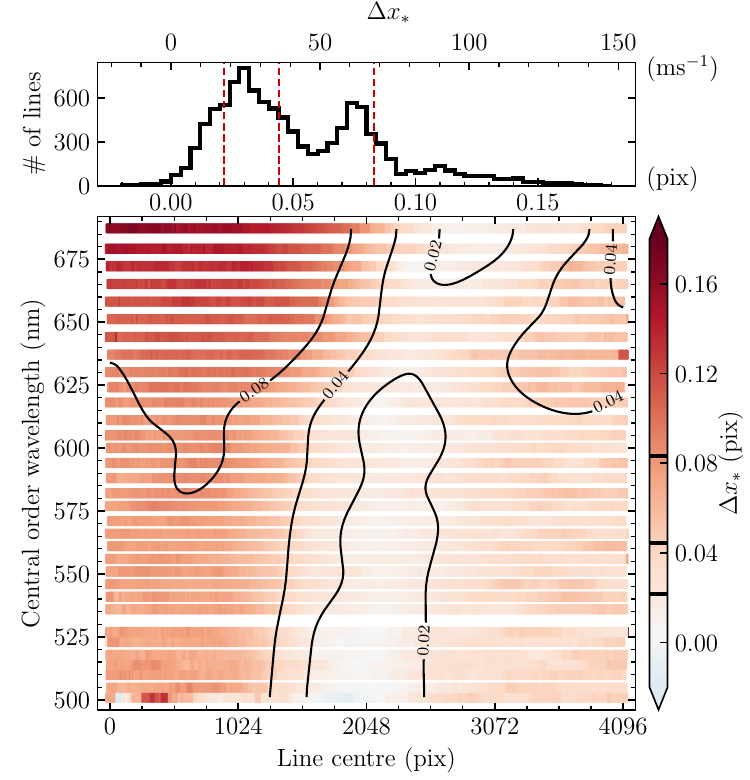}
    \caption{Differences between astrocomb line position measured by fitting two different IP models to the data, $\Delta x_* = x_{*,\rm G} - x_{*,\rm E} $. The subscripts indicate which IP model was used to determine the centre during line fitting (G for the Gaussian and E for our empirical IP models). Main panel: $\Delta x_*$ as a function of position on the detector. The zero point of the colour bar is set to $\Delta x_*= \pix{0}$, such that the red and blue colours correspond to positive and negative differences. Contours show the spatial distribution of values more clearly, with contour levels corresponding to the median and the central 68\% distribution limits. Top panel: The histogram of the values plotted in the main panel. The bottom $x$-axis is in units pixel and the top axis in units \mps{} ($\pix{1} = \mps{820}$). The vertical dashed red lines show the median and the central 68\% distribution limits. The same quantities are shown as horizontal thick black lines in the colour bar to the right of the main panel. }
    \label{fig:CCD_centre_differences}
\end{figure}

\begin{figure}
    \centering
    \includegraphics[width=\columnwidth]{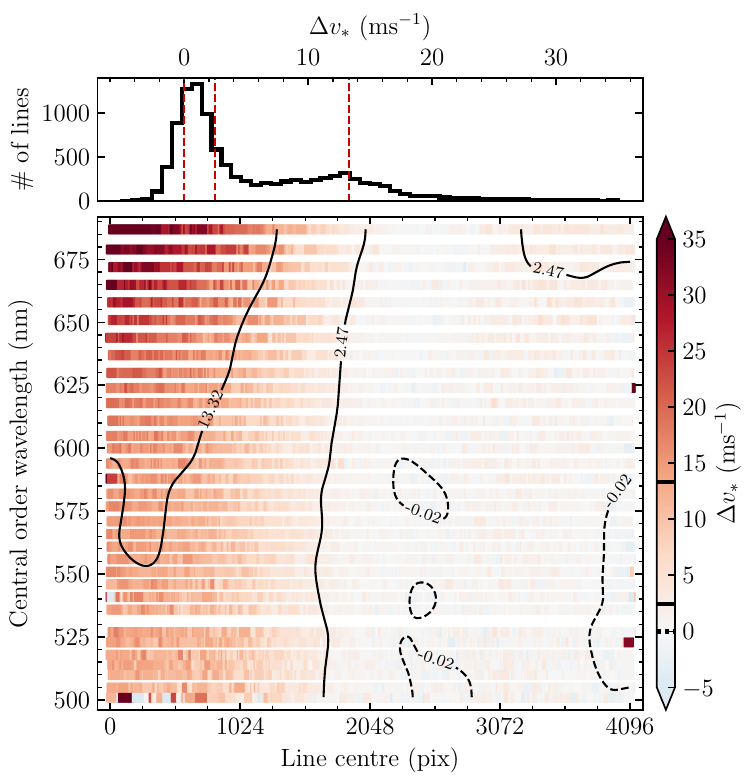}
    \caption{Differences between wavelengths of astrocomb lines measured using our empirical IP models and using Gaussian IPs, expressed as a velocity shift between the two (Eq.\ \eqref{eq:delta_v_*}). Main panel: $\Delta v_*$ as a function of position on the detector. The zero point of the colour bar is set to $\Delta v_* = \mps{0}$, such that the red and blue colours correspond to positive and negative velocity differences. Black contours more clearly visualise the distribution of $\Delta v_*$ values across the detector. Contour limits correspond to the median and the central 68\% distribution limits. A full line is used for positive contour levels and a dashed line for negative ones. Top panel: The histogram of the values plotted in the main panel. The vertical dashed red lines show the median and the central 68\% region limits. The same quantities are shown as horizontal thick black lines in the colour bar to the right of the main panel. 
    }
    \label{fig:CCD_lambda_differences}
\end{figure}

First we compared $x_*$ measured by fitting empirical IP shapes to $x_*$ determined by fitting Gaussian IP shapes. \figref{fig:CCD_centre_differences} shows $\Delta x_* = x_{*,\rm G}- x_{*, E}$ for all 10576 lines, revealing striking results. Firstly, zero differences are seldom observed. The maximum observed $\Delta x_* = \pix{0.193}$ (corresponding to a velocity shift of \mps{158}), appears at the blue wavelength edges of the reddest order (red area in the top left corner). Negative values are only ever observed in the very low S/N part of the spectrum (light blue area at the bottom centre). The 16\ts{th}, 50\ts{th}, and 84\ts{th} percentiles are \pix{0.020} (\mps{16.6}), \pix{0.039} (\mps{31.8}), and \pix{0.083} (\mps{67.7}), in that order. 50\% of the lines have $|\Delta x_{*}|<\pix{0.039}$ (\mps{31.8}) and 95\% of them have $|\Delta x_*|<\pix{0.116}$ (\mps{94.9}).

Secondly, the differences vary smoothly both within and across orders. For all orders, the maximum difference occurs at their blue wavelength edges (left in the main panel). Moving towards the red wavelength edge, $\Delta x_*$ approaches zero approximately in the middle of the order (coloured white) before increasing again at its red edge (turning more brightly red). The contours show this more clearly and also reveals that this pattern closely resembles the pattern for $\mu$ in \figref{fig:GP_hyperparameters_mosaic}.

Next, the two sets of $x_*$ measurements were used to wavelength calibrate HARPS. In producing the calibration, a single echelle order was divided into eight 512 pixel wide segments and a seventh order polynomial was fitted in each segment individually. No continuity conditions were imposed, meaning that the wavelength calibration is allowed to jump at each segment boundary. This approach was shown to be preferable over using a single continuous polynomial covering the entire order, even when very high order polynomials are used and detector stitching is accounted for \citep{Milakovic2020}. Both calibrations cover the wavelength range $\SI{498.9}{\nano\meter} \leqslant \lambda \leqslant \SI{691.4}{\nano\meter}$. Examining the histogram of the wavelength calibration residuals (not shown) for the two calibrations reveals they both approximately followed a normal distribution with zero mean and similar standard deviations. After outlier rejection, that is removing values falling more than $3\sigma$ away from the mean calculated over all 10576 values ($<1\%$ of the values), the standard deviation for the residuals of the calibration based on $x_{*,E}$ ($x_{*,G}$) was \mps{3.72} (\mps{3.82}). The residuals did not show any wavelength trends at any scale, except for the scatter being larger in lower S/N regions. Both calibrations are therefore ``correct'' in a statistical sense, meaning that the residuals could not be used to establish the accuracy of either calibration. 

The two wavelength calibrations were next compared to each other on a pixel-by-pixel basis. Expressing the differences in pixel wavelengths as velocity shifts, we found the same general pattern already seen in \figref{fig:CCD_centre_differences} so this is not shown. The only difference with respect to the pattern in \figref{fig:CCD_centre_differences} is that the pattern seen in wavelength calibration comparison is of the opposite sign. This can be understood by considering that, generally, $x_{*,E}<x_{*,G}$ so that the same wavelength (determined from Eq.\,\eqref{eq:lfc}) is associated with a smaller pixel number when $x_{*,E}$ is used. Once these centres are used for wavelength calibration, the same pixel is assigned a lower wavelength in the empirical IP case than in the Gaussian IP case, resulting in the velocity shift between the two calibrations having an opposite sign with respect to the sign of $\Delta x_*$. The median velocity shift between the two calibrations is \mps{-27.74} (Gaussian minus empirical IP calibration), with the central 68\% of the values falling between \mps{-76.04} and \mps{-16.06}. The maximum negative difference is approximately \mps{-150}, appearing at the same locations where the maximum positive $\Delta x_*$ is observed in \figref{fig:CCD_centre_differences}. This is consistent in amplitude with the simple conversion based on the velocity content of HARPS pixels.

\subsection{Differences between wavelength measurements made using Gaussian and empirical IP models}
To quantify the difference of our empirical IP models make on wavelength measurements from the data, the astrocomb spectrum was next treated as if it were a science spectrum, so it was calibrated using the wavelength calibration produced in \secref{sec:wavelength_calibration} and line wavelengths, $\lambda_*$, were measured from this spectrum without any prior knowledge from Eq.\,\eqref{eq:lfc}. 

For this purpose, we additionally created IP models in velocity space (shown in Appendix \ref{sec:IP_velocity}). The data and the procedures used for creating these models were exactly the same as described in Sections \ref{sec:methods} and \ref{sec:IP_reconstruction} (including the secondary GP for variance estimation), except that $\Delta x$ was expressed in terms of velocity instead of pixel. The velocity of the i\ts{th} pixel is:
\begin{equation} \label{eq:velocity_lambda_star}
    \frac{v_i}{c} = \frac{\lambda_i - \lambda_*}{\lambda_*} ,
\end{equation}
where $c$ is the speed of light, and $\lambda_i$ is the wavelength at pixel centre. Wavelengths for all pixels were obtained from the wavelength calibration, which was done independently in each iteration using the appropriate $x_*$ estimates. $\lambda_*$ is the wavelength at the astrocomb line centre determined by fitting our velocity space IP to the data in exactly the same way as was previously done in pixel space (\ref{sec:fitting_IP}), except for the weights in Eq.\,\eqref{eq:weights} changing to be $x_1=\kmps{2}$ and $x_2= \kmps{4}$. After a total of ten iterations, a set of most likely IP models in velocity space was saved separately and used here.

Finally, the wavelength of each line ($\lambda_*$) was measured from the wavelength calibrated data. Astrocomb lines were approximated by a Dirac $\delta$ function, for which the convolution with an IP model returns the IP model itself, so $\lambda_*$ was determined by fitting the appropriate IP shapes to the observations directly. Two sets of $\lambda_*$ measurements were produced: one by using our empirical IP shapes and the second by using Gaussians. To allow a more straightforward interpretation of the results, each kind of IP shape was used consistently to measure $x_*$ and $\lambda_*$. In both cases, the fitting followed the methods explained in \secref{sec:fitting_ip_to_data}, with $\lambda_*$ taking the place of $x_*$. Empirical IPs were fitted in velocity space because of the way they were constructed and because it was easier. Gaussian IPs were fitted in wavelength space instead. Doing so did not impact $\lambda_*$ measurements: the transformation from wavelength to velocity space results in the multiplication of the Gaussian function (Eq.\,\eqref{eq:gaussian_ip_fit_expression}) by $\exp{-(c/\lambda_*)^2}$, where $\lambda_*$ is the Gaussian centre and also defines zero velocity. This transformation can be written as a simple rescaling of Gaussian amplitude and width, but the centre remains unchanged. However, the transformation introduces correlations between the amplitude, the width and $\lambda_*$ so it was avoided. 

Differences between the two sets of $\lambda_*$ measurements were expressed in velocity units:
\begin{equation}\label{eq:delta_v_*}
    \frac{\Delta v_*}{c} = \frac{\lambda_{*,G}-\lambda_{*,E}}{\lambda_{\rm LFC}},
\end{equation}
where subscripts G and E on $\lambda_*$ again indicate the IP used and $ \lambda_{\rm LFC}$ is the expected wavelength of the line determined through Eq.\,\eqref{eq:lfc}. \figref{fig:CCD_lambda_differences} shows the results, with $\Delta v_*$ varying smoothly across the detector in a way similar to \figref{fig:CCD_centre_differences}. The largest positive (but also absolute) departures from zero can be found in the blue wavelength edges of the reddest echelle orders, reaching up to \mps{35}. Distribution median is \mps{2.47} and the central 68\% of the values are within $\mps{-0.02}\leqslant\Delta v_*\leqslant\mps{13.32}$. 16\% of the lines have $\Delta v_*$ below zero. In absolute terms, the 50\%, 70\%, and 95\% of the lines have $|\Delta v_*|\leqslant\mps{2.56}$, \mps{8.76}, and \mps{19.37}, respectively. 


While \figref{fig:CCD_centre_differences} shows the differences between $x_*$ when two different IP shapes are assumed, \figref{fig:CCD_lambda_differences} shows the difference between the end results of a chain of processes performed under two different assumptions for IP shape. These processes are, in order: measuring $x_*$, wavelength calibrating HARPS using $x_*$, constructing IP models in velocity space from wavelength calibrated data, and measuring $\lambda_*$ by fitting these models to the wavelength calibrated data. The fact that differences observed in \figref{fig:CCD_lambda_differences} are smaller than those in \figref{fig:CCD_centre_differences} demonstrate that wavelength calibration diminishes the initial differences between $x_*$ seen in \figref{fig:CCD_centre_differences}.

\begin{figure*}
    \centering
    \includegraphics[width=\columnwidth]{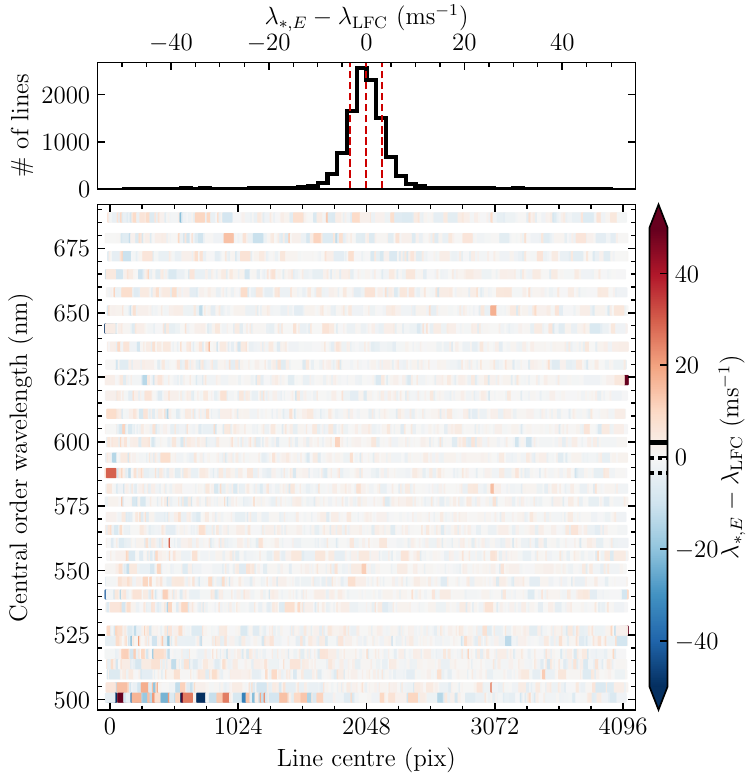}
    \hspace{0.2cm}
    \centering
    \includegraphics[width=\columnwidth]{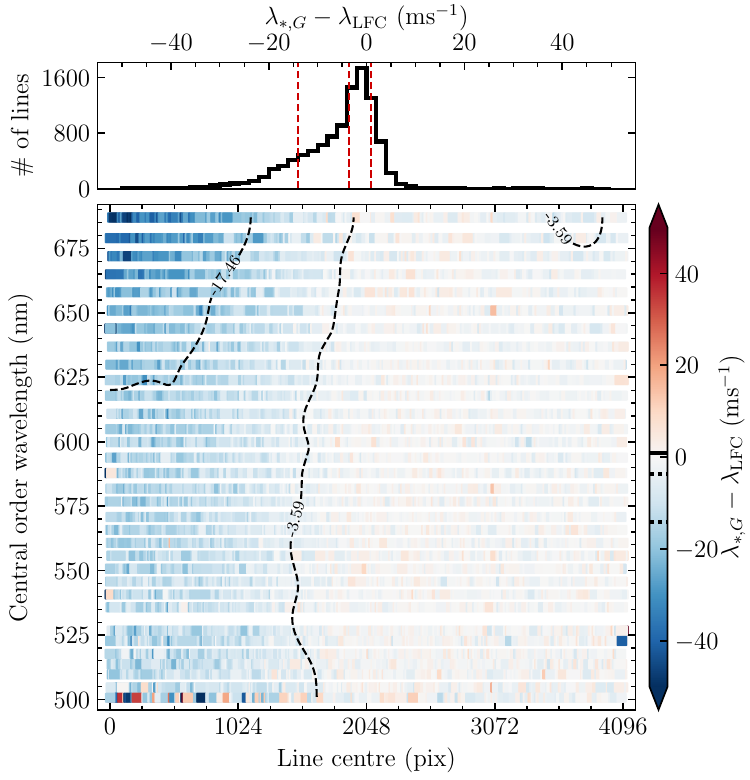}
    \caption{Wavelength measurement error, $\epsilon_\lambda$, as a function of detector position for our empirical IP models (left) and for the Gaussian IP  approximation (right). Colour schemes in both panels are identical and centred at \mps{0} (white) to highlight the vast improvements associated with the former. Contours for the empirical IP case (left) were not informative so are not shown. Plotted contour levels for the Gaussian IP case (right) correspond to the 16\ts{th} and 50\ts{th} value percentiles (\mps{-14.06} and \mps{-3.59}, respectively) while values above the 68\ts{th} percentile (i.e.\ above \mps{0.93}) do not cluster at any specific location so that contour is not shown. Smaller panels show the histogram of values shown under them. The median and the central 68\% distribution limits are shown as vertical dashed red lines. The same quantities are shown as thick black lines in the colour bar to the right of each main panel.}
    \label{fig:CCD_wavelength_errors}
\end{figure*}

The uncertainties on $\lambda_{*}$ associated with empirical IPs, $\sigma_{\lambda_{*,E}}$, resemble those on $x_{*,E}$ both in the distribution on the detector and in their histogram (so are not shown). The median value of $\sigma_{\lambda_{*,E}}$ is \mps{1.74}, with the 16\ts{th}, 68\ts{th}, and 95\ts{th} percentiles \mps{1.13}, \mps{2.75}, and \mps{4.09} (respectively). These numbers are in excellent agreement with the uncertainties on $x_{*,E}$ expressed in velocity units (c.f.\ with numbers in \secref{sec:fit_quality}). Such excellent agreement was not necessarily expected and gives further support to the validity of our methods. Values for $\sigma_{\lambda_{*,E}}$ are also consistent with theoretical predictions based on the photon counting statistics, $\sigma_{\rm pn}$ \citep{Bouchy2001}. The average $\sigma_{\rm pn}$ for a single line is \mps{2.39} and the 16\ts{th}, 50\ts{th}, 68\ts{th}, and 95\ts{th} percentiles of the $\sigma_{\rm pn}$ distribution over all lines are \mps{1.67}, \mps{2.00}, \mps{2.44}, and \mps{2.96} (in that order). The median ratio of $\sigma_{\lambda_{*,E}}/\sigma_{\rm pn}$ (calculated for each line individually) is 0.88 and the central 68\% of the ratios fall within 0.58 and 1.26.

\subsection{Wavelength scale distortions}
\label{sec:wavelength_distortions}

Another important question is whether choosing a particular IP produces correlated errors (distortions) in the wavelength scale of the instrument. Avoiding distortions is important whenever relative wavelength shifts between different spectral features are measured, as is done for fundamental constant studies and sometimes also for studies of isotopic abundances of chemical elements. The existence of wavelength scale distortions in UVES \citep{Dekker2000}, HIRES \citep{Vogt1994}, and HDS \citep{Noguchi2002} spectrographs is well documented and their impact on fundamental constant measurements is understood \citep{Rahmani2013,Evans2014,Whitmore2015,Dumont2017,Milakovic2020}. In fact, removing wavelength scale distortions is one of the major expected benefits from using astrocombs for instrument wavelength calibration and a major reason for their deployment on current and future spectrographs. Yet, previous studies found some small level of distortions can remain even when astrocombs are used \citep{Milakovic2020,Schmidt2021}.

\begin{figure*}
    \centering
    \includegraphics[width=\textwidth]{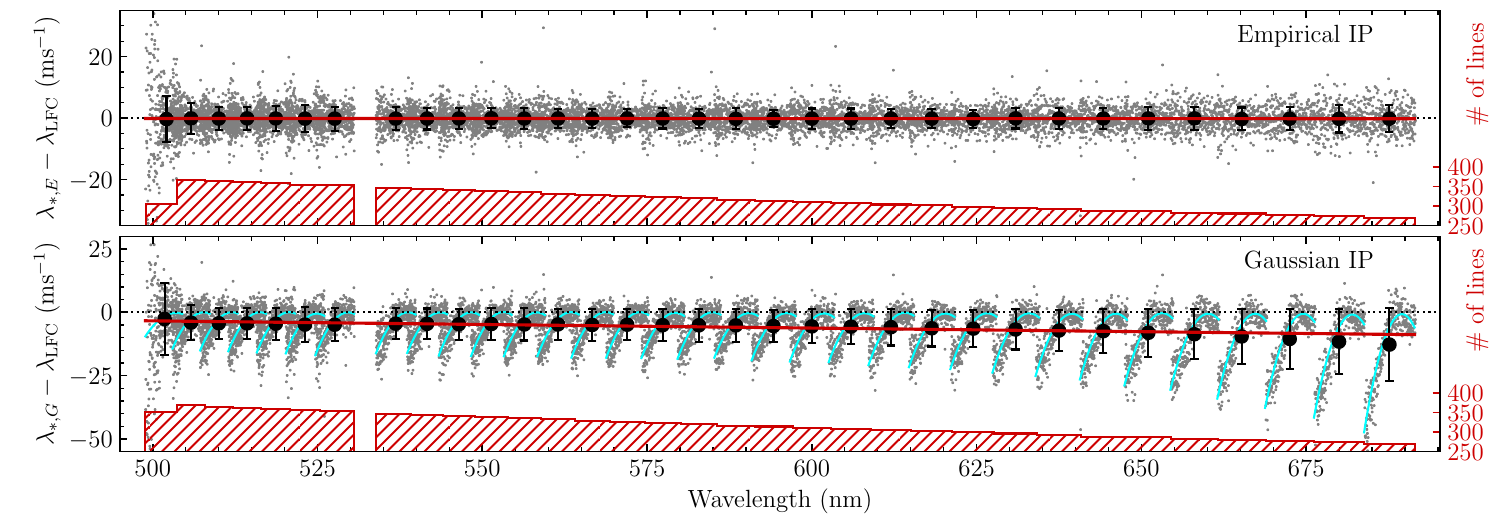}
    \caption{Wavelength measurement error, $\epsilon_\lambda$, as a function of wavelength for our empirical IP models (top) and for the Gaussian IP approximation (bottom). Grey dots are the same 10576 error values shown in \figref{fig:CCD_wavelength_errors}, but now plotted as a function of $\lambda_{\rm LFC}$. The data were binned such that each bin contains all astrocomb lines appearing in the same echelle order, after which the mean and standard deviation were calculated for each bin (black dots with error bars). In calculating the above quantities, we ignored values falling outside of $\pm\mps{20}$ ($\pm\mps{50}$) for the empirical (Gaussian) IP cases. The red line in each panel is the best-fit straight line going through the binned values (see text for parameter values). The red hashed histogram and the red labels on the right spine of the Figure show the number of grey dots falling in each bin. Note that the bin edges are approximate because of the partial overlap between spectral ranges of adjacent orders. Thin cyan lines in the bottom panel are the best fit second order polynomials to the grey dots. Their parameters are tabulated in \tabref{tab:distortion_coefficients}}
    \label{fig:lambda_error_vs_wave}
\end{figure*}

Errors on $\lambda_{*}$ measurements, i.e. $\epsilon_\lambda = \lambda_{*}-\lambda_{\rm LFC}$, are shown in \figref{fig:CCD_wavelength_errors} for all 10576 lines, expressed in \mps{}. The empirical IP case (left) shows no evidence of wavelength scale distortions at any scale and no dependence on position on the detector. The larger scatter at in the order with a \nm{500} central wavelength below \pix{2048} is due to low S/N in data. The histogram of the errors (top panel, left) approximately follows a normal distribution centred at zero. After removing 78 lines (0.7\% of the total) with errors falling outside of the $\pm\mps{20}$ range, the remaining 10498 error values have a mean of \mps{-0.17} and a standard deviation of \mps{3.75}. The latter value is comparable to the standard deviation of the wavelength calibration residuals (\mps{3.72}) but is $\approx60\%$ larger than the average $\sigma_{\rm pn}$ (\mps{2.39}) and the average $\sigma_{\lambda_{*,E}}$ (\mps{2.30}). The median error is \mps{-0.21}, and the central 68\% of the error values are in the \SIrange{-3.47}{3.18}{\meter\per\second} range. 95\% of the lines have absolute errors smaller than \mps{7.54}.

The right side of \figref{fig:CCD_wavelength_errors} shows $\epsilon_\lambda$ when the Gaussian IP approximation is used instead. In this case, the histogram of the errors (top panel) shows a strong tail extending towards \mps{-50}. 30 values (0.3\% of the total) falling outside of $\pm\mps{50}$ range were removed from further consideration. The median of the remaining values is \mps{-3.59} and the central 68\% of the values fall within $\mps{-14.06}$ and $\mps{0.93}$. Positioning the errors on the detector (main panel), clear evidence of wavelength scale distortions is revealed. The distortions appear both in the main dispersion direction (horizontally) and in the cross-dispersion direction (vertically), with the detector divided into approximately equal halves vertically following the 50\ts{th} percentile contour level. This division is consistent with the patterns already seen in Figs.\,\ref{fig:CCD_centroid-gaussian}, \ref{fig:CCD_centre_differences} and \ref{fig:CCD_lambda_differences}. The largest errors generally appear at the blue order edges of orders and the largest absolute errors are in the reddest echelle orders (see the contour tracing the 16\ts{th} percentile in the same panel).

We further examined the long-range and short-range distortions by plotting $\epsilon_\lambda$ as a function of wavelength, shown as small grey dots in \figref{fig:lambda_error_vs_wave}. To better examine long-range trends, $\epsilon_\lambda$ values were binned according to the echelle order that they appear in, after which the mean and the variance of the values in the each bin were calculated (large black dots with error bars in the same Figure). In calculating the means and the variances, we removed values falling outside of $\pm\mps{20}$ ($\pm\mps{50}$) for the empirical IP (Gaussian IP) case. The red hashed histograms in the two panels of \figref{fig:lambda_error_vs_wave} show how many values remained after outlier removal. At the end, a straight line of the form 
     $ m (\lambda-\lambda_0) + b$
was fitted to the mean errors with inverse variances as weights, and the result is shown as a red line in each panel. For $\lambda_0=\nm{600}$, the top panel of \figref{fig:lambda_error_vs_wave} has $m=\SI{-0.04(2)}{\meter\per\second}$ per \nm{100} and $b=\mps{-0.17(1)}$ and $m=\mps{-2.8(3)}$ per \SI{100}{\nano\meter} and $b=\mps{-6.2(2)}$ for the bottom panel. 

We found no evidence for short-range distortions associated with the empirical IP case (grey dots in the top panel of \figref{fig:lambda_error_vs_wave}). Concentrating on the Gaussian IPs case, that is on the grey dots in the bottom panel of \figref{fig:lambda_error_vs_wave}, we observed strong inter-order distortions. In the worst case, the total distortion within one order can be as large as \mps{60}. The distortion within a single echelle order seemed to be well described by a second order polynomial (cyan lines going through the grey dots in the bottom panel). Their best fit coefficients are reported in \tabref{tab:distortion_coefficients}. 

\subsection{Examination of astrocomb modes appearing twice on the detector}
\label{sec:repeating_modes}
Some astrocomb modes appear twice on the detector due to the spectral range overlap between two adjacent orders, offering another powerful tool to examine the wavelength measurement accuracy and the existence of any biases. If the IP models used to fit astrocomb lines are correct and if the wavelength calibration derived from astrocomb centres is accurate, the measured $\lambda_*$ for the two observations of the same mode should be identical. We identified 1472 such modes in the spectrum and calculated the relative shift between the pair of wavelength measurements:
\begin{equation}\label{eq:Delta_v_pair}
    \frac{\Delta v_{\rm pair}}{c} = \frac{\lambda_{*,2}-\lambda_{*,1}}{\lambda_{\rm LFC}}.
\end{equation}
In the formula above, $\lambda_{*,1}$ and $\lambda_{*,2}$ are the two measured wavelengths for the same mode appearing at two locations of the detector, and $c$ is the speed of light.

\begin{figure}
    \centering
    \includegraphics[width=\columnwidth]{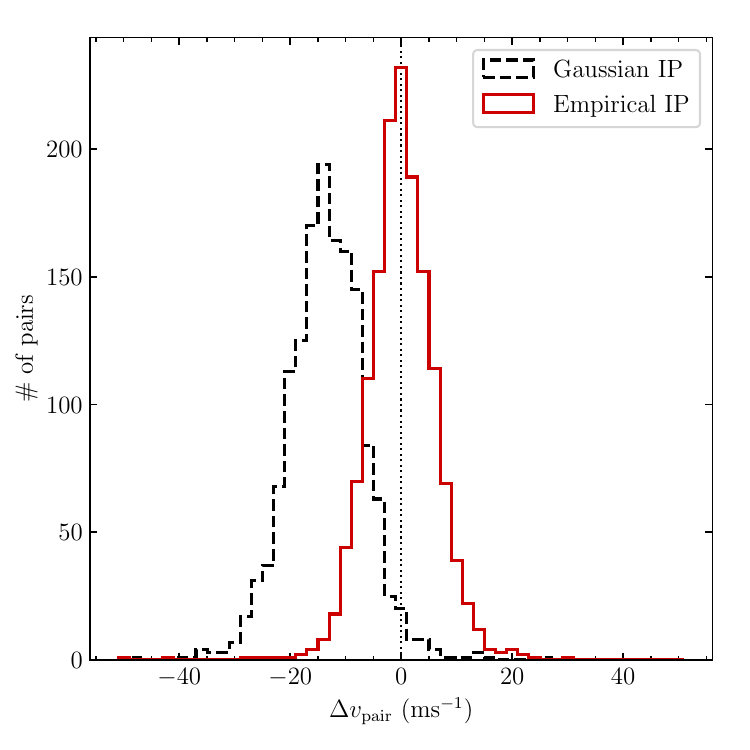}
    \caption{Histogram of measured velocity shifts (Eq.\,\eqref{eq:Delta_v_pair}) between 1472 astrocomb modes appearing twice in the spectrum. Line wavelengths were measured by fitting the most likely empirical IP model (full red histogram) or a Gaussian IP model (dashed black histogram) to the data in a wavelength calibrated spectrum. Zero velocity means that the same wavelength was measured for both lines of the pair.}
    \label{fig:wavecal_repeated_lines_histogram}
\end{figure}

The distribution of $\Delta v_{\rm pair}$ associated with the empirical IP case is plotted as the red histogram in \figref{fig:wavecal_repeated_lines_histogram} and shows that here is no bias associated with the use of our empirical IP models. In comparison, using Gaussian IPs results in a strong bias (dashed black histogram in \figref{fig:wavecal_repeated_lines_histogram}). The statistical properties of these distributions can be found in \tabref{tab:repeated_lines_stat}. Interpreted as probability distributions, the histograms can be used to quantify accuracy associated with the use of a specific IP shape. For the empirical IP case, the probability of achieving accuracy better than \mps{20}, \mps{10}, and \mps{5}is 99.9\%, 92.0\%, and 64.1\%. The same levels of accuracy for the Gaussian IP case have probabilities of 84.3\%, 28.8\%, and 8.4\% (respectively). 

\begin{figure}
    \centering
    \includegraphics[width=\columnwidth]{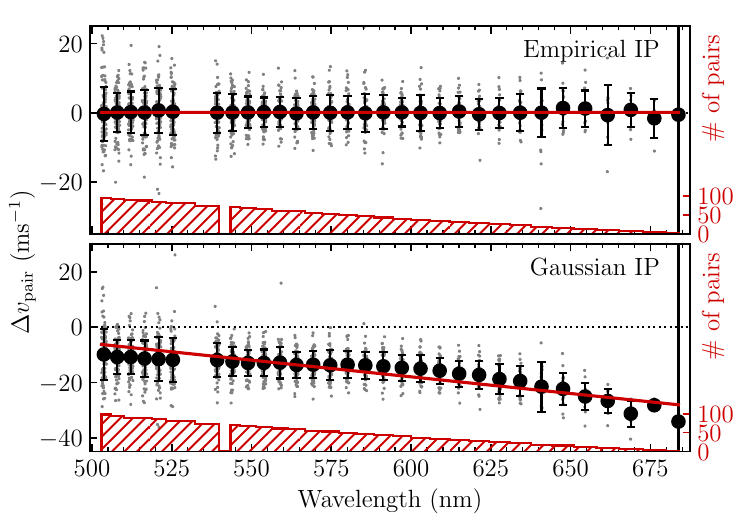}
    \caption{Dependence of $\Delta v_{\rm pair}$ on wavelength when our empirical IPs are used (top panel) and when Gaussian IPs are used (bottom panel). Small grey dots are the 1472 $\Delta v_{\rm pair}$ values and large black dots with error bars are the means and the standard deviations calculated over binned values. Bin edges were determined such that a single bin contain all astrocomb lines appearing in the same echelle order. In calculating the above quantities, we ignored values falling outside of $\pm\mps{20}$ ($\pm\mps{50}$) for the empirical (Gaussian) IP cases. The red line in each panel is the best-fit straight line going through the binned values (see text for parameter values). The red hashed histogram and the red labels on the right spine of the Figure show the number of grey dots falling in each bin. The bin at $\lambda=\nm{687}$ contains a single point. }
    \label{fig:accuracy_vs_wavelength}
\end{figure}
\figref{fig:accuracy_vs_wavelength} shows $\Delta v_{\rm pair}$ as a function of wavelength, analogously to \figref{fig:lambda_error_vs_wave}. Grey dots show individual values and black dots with error bars show the mean and the standard deviation in bins, where each bin contains all lines falling in the same echelle order. The empirical IP case shows no wavelength dependency at any scale (top panel). The best fitting straight line going through the binned $\Delta v_{\rm pair}$ values in the top panel has a slope $m=\mps{0.0(2)}$ per 
\SI{100}{\nano\meter} and offset $b=\mps{0.1(1)}$ at $\lambda_0=\nm{600}$. The Gaussian IP case shows a long-range linear trend of $\Delta v_{\rm pair}$ with wavelength, with $m=\mps{-12.1(7)}$ per \SI{100}{\nano\meter} and $b=\mps{-18.0(4)}$. At the two extremes, the observed $\Delta v_{\rm pair}$ is \mps{-5.9} (at $\lambda=\nm{500}$) and \mps{-28.9} (at $\lambda=\mps{690}$). Extrapolating the straight line outside of the fitted range, $\Delta v_{\rm pair}=
\mps{0}$ is expected at $\lambda=\nm{451}$ and the blue end of the HARPS wavelength range ($\lambda=\mps{380}$) should have $\Delta v_{\rm pair} = \mps{8.6}$. Although a straight line explains the data sufficiently well, the smooth variation of the binned values with wavelength may hint towards a more complex, non-linear, trend. 

\begin{table}
\centering
    \caption{Statistical properties of the distributions shown in \figref{fig:wavecal_repeated_lines_histogram}.  
   }
    
    \begin{tabular}{l S[table-format=3.2] S[table-format=3.2] }
        \hline\hline
        {Quantity} & {\multirow{2}{*}{Empirical IP}} & {\multirow{2}{*}{Gaussian IP}} \\
                    {($\mps{}$)} &  & \\
        \hline
        Mean         &  -0.16 & -13.67 \\
        Std.\,dev.   &  8.24  & 8.08 \\
        Median       &  -0.05 & -13.48 \\
        MAD\tablefootmark{\textdagger}          &  3.59  &  4.34 \\
        \multicolumn{3}{l}{Percentiles:}   \\
        \multicolumn{1}{r}{16\ts{th}}  &   -5.52      &   -19.97    \\
        \multicolumn{1}{r}{84\ts{th}}  &    5.65      &    -7.18    \\
        \multicolumn{1}{r}{98\ts{th}}  &   12.91      &     0.67    \\
  \hline
    \end{tabular}
    \tablefoot{
    \tablefoottext{\textdagger}{Median absolute deviation (around the median).}
    }
    \label{tab:repeated_lines_stat}
\end{table}

%
\section{Results}\label{sec:result_summary}
%
We conducted a detailed study of the HARPS IP using its astrocomb calibration system. Applying the methods described in \secref{sec:methods} to a single astrocomb calibration frame taken in December 2018, we reconstructed the IP in the wavelength range \SIrange{500}{690}{\nano\meter} (\secref{sec:IP_reconstruction}). 

The main results of this work are:
\begin{enumerate}
    \item HARPS IP significantly differs from a Gaussian shape in the entire range covered by the astrocomb (Figs.\, \ref{fig:stack_illustration}, \ref{fig:CCD_centroid-gaussian}, \ref{fig:IP_example}). The IP asymmetry varies quickly across the detector in both the main dispersion direction and in the cross-dispersion direction (\figref{fig:IP_all_pixel}). In every echelle order illuminated by the astrocomb, the IP exhibits a rightwards tail that is the strongest at the blue wavelength edge and gradually diminishes with increasing wavelength. Tail strength varies also in the cross-dispersion direction (i.e.\ across different orders), with the reddest orders having the strongest tails (Figs.\,\ref{fig:IP_all_pixel} and \ref{fig:IP_all_velocity}). 

    \item Assumptions on IP shape impacts significantly on the instrument's wavelength calibration and on wavelength measurement accuracy. Centres of astrocomb lines measured by fitting Gaussian shapes differed by as much as \pix{0.193} (\mps{158}) from their centres measured by fitting empirical IP shapes (\figref{fig:CCD_centre_differences}). The median difference between the two centre estimates over all 10576 lines was \pix{0.039} or \mps{32}. When these centres were used to wavelength calibrate HARPS, the differences between the two wavelength calibrations were of similar magnitudes with an opposite sign (median shift was \mps{-27.74} and maximal shift was \mps{-150}). 

    \item Using the Gaussian IP assumption is associated with errors on wavelength measurements as large as \mps{50} (\figref{fig:CCD_wavelength_errors}, right). The errors reveal both intra-order and inter-order wavelength scale distortions which are generally larger at locations where the empirical IP shapes are more strongly asymmetric. The intra-order distortions associated with Gaussian IP use are described well by a second order polynomial and the long-range inter-order distortion by a linear function with a slope $m=\mps{-2.8(3)}$ per \SI{100}{\nano\meter} and a normalisation $b=\mps{-6.2(2)}$ at $\lambda_0=\nm{600}$ (\figref{fig:lambda_error_vs_wave}, bottom). 

    \item No bias was associated with the use of our empirical IP models. Errors on wavelength measurements for 10576 astrocomb lines using empirical IP models follow an approximately normal distribution with a \mps{-0.21} median and \mps{2.21} median absolute deviation. No correlations of errors with position on the detector (\figref{fig:CCD_wavelength_errors}, left) or with wavelength (\figref{fig:lambda_error_vs_wave}, top) were observed. Examination of 1472 modes appearing twice on the detector (\secref{sec:repeating_modes}) confirmed this further. Using empirical IP models, wavelength accuracy was better than \mps{10} (\mps{5}) for 92\% (64\%) of the lines. For comparison, only 29\% (8\%) of lines were measured with the same accuracy using the Gaussian IP. 

    \item Distortions associated with the Gaussian IP assumption change smoothly across the detector area, suggesting a connection to spectrograph optics. Similar distortions may therefore be seen in other cross-dispersed spectrographs when the Gaussian assumption is used for wavelength calibration.

\end{enumerate}
%
\section{Discussion }\label{sec:discussion}
%
Astrocombs are widely believed to be the ideal absolute wavelength calibration sources, providing accuracy and precision that is limited only by photon counting statistics. This is why several existing and planned high-resolution spectrographs are (expected to be) equipped with astrocombs for wavelength calibration and instrument monitoring and characterisation. While the first statement is likely true, this work demonstrated that the knowledge of the spectrograph's instrument profile is crucial in unlocking the full potential of astrocombs. By modelling its empirical IP, we achieved a large step forward in terms of achieving accurate wavelength measurements with HARPS. 

The IP reconstruction made use of the flexible nature of GP regression, an advantage independently recognised by \citet{Hirano2020}, who first used it to reconstruct the IP of a high-resolution astronomical spectrograph. While both we and \citet{Hirano2020} modelled the IP in small sections of the detector, the details of this process differed between us in one major way: we normalised the astrocomb lines and stacked all lines falling in one segment on top of their centres before performing the regressions whereas \citet{Hirano2020} fitted the fluxes of all lines in the segment simultaneously. Another approach to IP reconstruction was put forward by \citet{Hao2020}, who used modified Gaussian functions to determine the dominant, smoothly changing IP shape, and used cubic splines to model residuals of this dominant component. The inclusion of a Gaussian mean function in our GP (\secref{sec:methods}) renders the two approaches similar. The normalised residuals shown in their figure A.1 bear close resemblance to $\psi - \mathbf{m}$ for our models. One such example is shown in \figref{fig:IP_example} as a dashed magenta line and in the second panel of \figref{fig:IP_example_detailed_scatter=True}. Unfortunately, we could not compare our IP models to the previous investigation of the HARPS IP using astrocombs by \citet{Zhao2021} because no IP models were produced as a part of that work. 

Our IP models are described by set of 10 hyperparameters (\secref{sec:methods}). Interestingly, albeit they were derived completely independently, the values of some hyperparameters vary smoothly across the detector area (\figref{fig:GP_hyperparameters_mosaic}), suggesting that they trace changes in some physical quantities. This is most obviously seen for the hyperparameters describing the mean function ($A$, $\mu$, $\tau$, and $y_0$), but can also be seen for $a$, $l$, and possibly other hyperparameters. Choosing a more physically motivated set of hyperparameters, based on considerations of instrument optics, or constraining them to smoothly change across the detector area similarly to \citet{Hao2020} may provide improvements in the future. 

We found conclusive evidence that assuming a Gaussian IP shape introduces both intra-order and inter-order wavelength scale distortions(Secs.\,\ref{sec:wavelength_distortions} and \ref{sec:repeating_modes}). These distortions range between \SIrange{5}{50}{\meter\per\second} and vary smoothly across the detector area (\figref{fig:CCD_wavelength_errors}). The latter suggest a connection with the optical design of HARPS. As such, similar distortions may appear in the wavelength calibrations of other cross-dispersed echelle spectrographs when their IP shape is assumed to be Gaussian. 

Left uncorrected, these distortions negatively impact on some types of analyses made from the data calibrated using the Gaussian approximation. Their severity, of course, depends on where the spectral features used in the analysis fall with respect to the distortion and the specifics of the analysis being performed. The impact on $\daoa$ measurements from quasar absorption systems may be important. For example, a \mps{50} distortion between \ion{Fe}{ii} $\lambda2382$ and \ion{Mg}{ii} $\lambda$2796 (commonly used for $\alpha$ measurements), could be erroneously interpreted as evidence for $\daoa=\SI{3e-6}{}$. 

Spatial variations of the IP across the detector also affect velocity measurements via the Doppler effect. The spectra of objects (quasars or stars in the case of exoplanet studies) shift by $\pm$ \SI{30}{\kilo\meter\per\second} within 1 year due to changes in the Barycentric Earth Radial Velocity, corresponding to a maximal difference of $\approx80$ pixels on HARPS. Measuring line centres using a Gaussian IP when the correct IP is, in fact, significantly asymmetric (such as it is in the top left angle of the right panel of \figref{fig:CCD_wavelength_errors}), is expected to bias the result by $\sim\mps{10}$. This is diminished by averaging over other lines used to measure the object velocity, with several thousands being a typical number used in exoplanet studies and the expected number of Lyman-$\alpha$ lines to be used for the redshift drift measurement \citep{Liske2008}. As an order of magnitude approximation, when averaging over 3000 lines, a \mps{10} shift in a single line would result in a measured global shift of \cmps{18} for the entire spectrum. This is similar to the expected signal from an Earth-Sun analogue ($\pm\cmps{9}$ in a year) and $60\times$ larger than the cosmological signal expected from redshift drift in $\Lambda$CDM ($\cmps{0.3}$ per year). Should Gaussian IP approximation be used in such studies, a full simulation (considering the locations at which the lines appear on the detector) should be run to quantify instrumental uncertainties. 

The presence of intra-order wavelength distortions raises important questions about the practice of order merging when combining object spectra as merging orders containing uncorrected distortions creates a complicated wavelength scale distortion in the combined spectrum. It may also smear out spectral features falling in the affected regions, complicating analysis. A quick (but possibly not very accurate) correction of intra-order distortions in wavelength calibrations produced by fitting Gaussian profiles to HARPS astrocomb spectra, is provided by the coefficients in Appendix \ref{app:distortion_coefficients}. We warn that we did not check whether the distortions are stable in time so these coefficients may not be applicable for data not taken shortly before or after 2018-12-07. 

In \secref{sec:background} we presented our method for background modelling. To focus on our main aim of IP reconstruction, we did not elaborate discussion of the background further. However, we note that procedures relating to the spectral background determination had a noticeable impact on the final fit quality for astrocomb lines. The background is highly modulated and, looking at its power spectrum, we saw that it contains significant power at periods as short as \pix{30}. Significant effort was therefore invested into determining the locations in the minima between the astrocomb modes as accurately as possible and into the background interpolation methods. Future attempts to model the IP from astrocomb spectra could investigate whether further improvements can be made by relating the modulation of the spectral background (connecting minima) to the modulation of the spectral envelope (connecting maxima). Furthermore, astrocombs based on technologies which do not require a high-power amplification stage (e.g.\ opto-electric LFCs, \citet{Zhuang2023}) may have smaller (or negligible) background levels so should be less sensitive to this issue.

Tables containing the numerical values describing empirical HARPS IP shapes are provided in the supplementary material in a machine readable format and also as a dataset on the following link: \url{https://zenodo.org/doi/10.5281/zenodo.10492989}. We hope that their use leads to interesting scientific discoveries.

\section{Conclusions \& future prospects}\label{sec:conclusions}

With some modification, our (or other similar) methods can be applied to any spectrograph equipped with an astrocomb, such as ESPRESSO \citep{Pepe2021}, EXPRES \citep{Blackman2020}, and Veloce \citep{Gilbert2018}. In the future, astrocombs will also be installed on the Keck Planet Finder \citep{Wu2022}, and two spectrographs to be mounted on 30-m class telescopes: ANDES \citep{Marconi2022} and G-CLEF \citep{Szentgyorgyi2018}, and possibly also on novel and very interesting spectrograph designs fed by an array of smaller telescopes \citep{Eikenberry2019,Angel2022}. Most demanding projects planned with these instruments rely on these instruments, not only reaching wavelength calibration stability at the few \cmps{} level, but also maintaining it over periods of years or decades. This presents several large technical and methodological challenges that are yet to be resolved, such as resolving the \cmps{60} zero-point offset in the wavelength calibration when one astrocomb is replaced by another \citep{Milakovic2020}. The impact of our empirical IP models on that result will be reported in a separate publication. 

In the future, an astrocomb with tunable $f_0$ and/or tunable $f_r$ \citep[see][e.g.]{Savchenkov2008PhRvL.101i3902S,Yan2023ApOpt..62.6835Y} could be used to determine the IP with unprecedented detail. Changing $f$ by $\Delta f \sim \SI{100}{\mega\hertz}$ would shift the astrocomb line centres by 5-10\% of the pixel, allowing for the reconstruction of the IP locally without the need for stacking neighbouring lines or interpolation across large fractions of the detector area. Covering the full frequency range between lines, $f_r$, would then map out the IP at every pixel, as well as allow for the reconstruction of the intra-pixel response function $\mathcal{Q}$. 

A comparison with the IP derived from an \iodine~absorption cell would also be interesting. \iodine~cells have previously been used to determine the IP shape of optical echelle spectrographs directly from the science observations \citep{Marcy1992,Butler1996}. HARPS was equipped with an \iodine~absorption cell but it has been decommissioned in 2004 \citep{HARPS_manual}, meaning that a direct comparison of the IP derived from astrocomb calibrations and \iodine~cell calibrations was not possible on HARPS. There is an ongoing project to compare the IP derived from astrocomb against the IP derived from \iodine~absorption cell for ESPRESSO as a part of preparations for ANDES. 

Going from modelling the 1-dimensional IP (the line-spread function, as done here) to modelling the 2-dimensional IP (that is the point-spread function, PSF) would be very useful for improving spectral extraction procedures. Firstly, information on the PSF shape could be used to investigate and measure the effects of charge transfer efficiency and the intra-pixel sensitivity function by monitoring shifts in the centres of individual astrocomb lines when flux levels or astrocomb frequencies are varied in a controlled way. These effects could then be modelled and their impact removed from the raw data before spectral extraction. Flux dependent shifts of the HARPS LFC in 2d raw images was already explored in \citet{Zhao2021}, who found that they can be equivalent to radial velocity shifts of \mps{15}. Secondly, the PSF shapes and the 2d wavelength calibration of raw frames would produce important calibration information useful for advanced spectral extraction algorithms such as ``spectro-perfectionism'' \citep{Bolton2010}. 

Should all of the above be implemented to several high-resolution spectrographs equipped with astrocombs, this would provide the ability to directly compare and even combine spectra obtained using different spectrographs and different telescopes with \cmps{1} accuracy. Currently, the spectra of the same objects obtained using different instruments are often not compatible, complicating result interpretation. The most obvious example relates to the studies of quasar absorption systems in the context of fundamental coupling constants' variability \citep{Webb1999,King2012}. For example: the physical model parameters of the absorption system at $z=1.15$ towards the quasar HE0515$-$4414 are different when derived from UVES, HARPS, or ESPRESSO observations \citep{Kotus2017,Milakovic2021,Murphy2022}.

Combining data from different instruments and telescopes may be necessary to detect Earth-Sun analogues via the radial velocity (RV) method. Comparing RV measurements of Sun as a star made by four extremely stable spectrographs, \citet{Zhao2023AJ....166..173Z} finds that day-to-day variations in the mean daily RV measured by the four instruments agree within $\approx \cmps{50}$ (see their figure 9). It would be interesting to see whether this number decreases when IP variations are considered in RV calculations, although other systematic effects considered by the authors may be dominant.


\section*{Acknowledgements}
D.\,Milakovi{\'c} thanks Luca Pasquini, John Webb, and Paolo Molaro on discussions which improved this work. This research was funded in part by the Austrian Science Fund (FWF) SFB 10.55776/F68 ``Tomography Across the Scales'', project F6811-N36 (Advancing Extragalactic Archeology through Novel Inversion Techniques). This work made use of Python programming language packages \texttt{numpy} \citep{Harris2020_numpy}, \texttt{scipy} \citep{2020SciPy-NMeth}, \texttt{matplotlib} \citep{Hunter2007_matplotlib}, and \texttt{fitsio} \citep{fitsio}, in addition to the software already cited. We also thank the referee for useful suggestions.



\bibliographystyle{aa} 
\bibliography{bibliography} 




%
\begin{appendix}

\section{Choosing a centre for an asymmetric profile}\label{app:centre_estimator}
Our method requires that some measure of an IP centre is provided, so that line centres can be determined through fitting (Eq.\, 7-9). For asymmetric profiles, there is no meaningful definition of the line centre and, looking through the relevant literature, we found scarce information about this particular point. 

\ak~introduced their own centre estimator based on the considerations for the distribution of flux on the detector produced by an unresolved source. They posited that the IP is properly centred when an unresolved source located at the boundary between the two brightest pixels results in equal fluxes in the boundary pixels, i.e. $\psi(\Delta x  = -0.5)=\psi(\Delta x = 0.5)$. As a reminder, $\Delta x$ is the distance of the pixel centre from the unresolved spectral feature. This definition is of the centre depends, of course, on the IP shape in the central region but its only assumption is that the IP is a smooth function peaking somewhere in between the two pixel centres. It was developed specifically to deal with asymmetric IP shapes but its performance with respect to other centre estimators has never before been quantified so we were unsure whether it was the most appropriate for our purposes.

\begin{figure}
    \centering
    \includegraphics[width=\columnwidth]{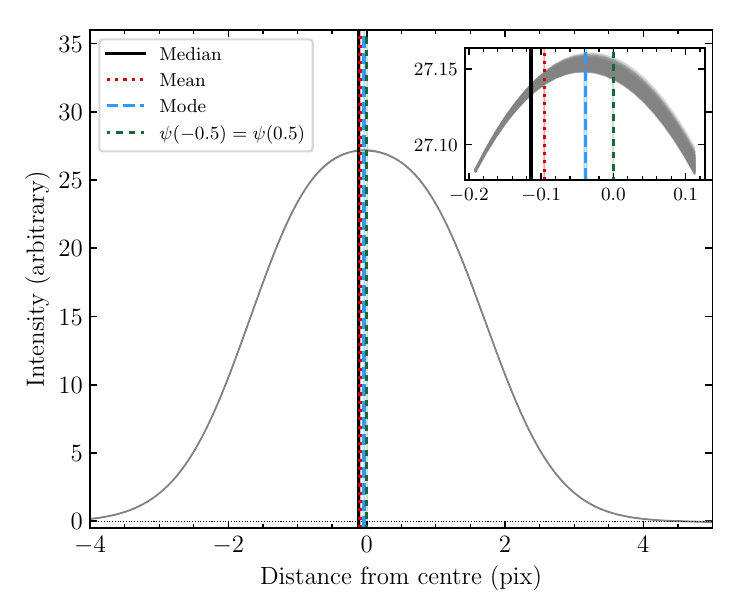}
    \caption{Set of 1000 MCMC HARPS instrumental profiles, trained on the data and shown on a fine grid. For each profile, we calculated the median, the mean, the mode, and the location satisfying $\psi(-0.5)=\psi(0.5)$, producing distributions for each of the four quantities. The means of the four distributions is shown as a vertical line of different colour and stroke. The $x$-axis zero point is set to the mean of the distribution for $\psi(-0.5)=\psi(0.5)$ centres. The inset on the top right shows a zoom-in of the region around the profile peak. Units in the inset are the same as in the main panel. }
    \label{fig:centre_comparison_ip_model}
\end{figure}

We therefore performed a numerical exercise to determine what is the most appropriate centre estimator for asymmetric IPs. The criterion chosen for their ranking was the estimator's invariability under small perturbations of the IP shape (robustness), where the best estimator would be the least variable. Four different estimators were considered: the mean, the median, the mode, and the definition of line centre from \citet{Anderson2000}. The numerical exercise made use of the MCMC models from Appendix \ref{app:mcmc} and posterior distributions for hyperparameter values in particular. The posterior distribution for relevant hyperparameters was sampled 1000 values to create 1000 IP models shown in \figref{fig:centre_comparison_ip_model}. Centres of these models were subsequently determined using each of the four estimators and their distributions were examined.

\begin{figure}
    \centering
    \includegraphics[width=\columnwidth]{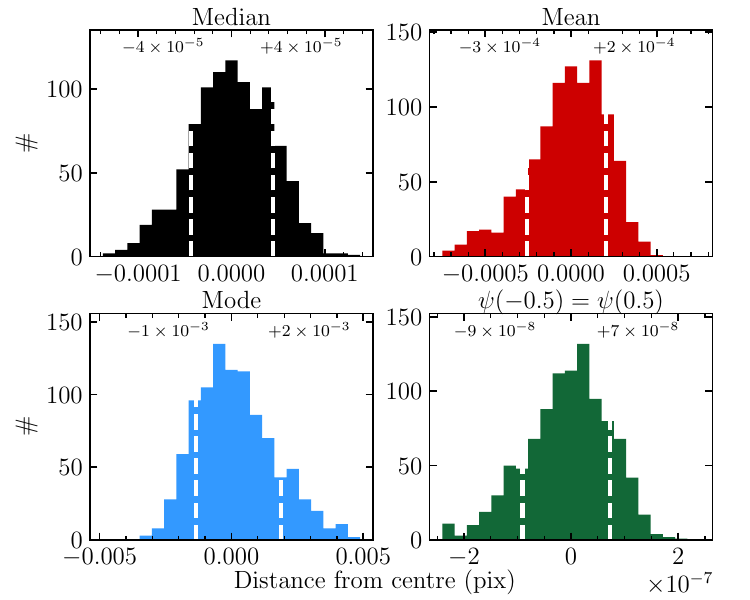}
    \caption{ Distribution of `centre' measurements for the 1000 instrumental profiles shown in \figref{fig:centre_comparison_ip_model}, where the centre estimator used is indicated at the top of the panel. Since we were interested in quantifying the stability (distribution width) of each centre estimator under small perturbations of the IP, the distributions plotted were shifted to have zero median. The dashed white vertical lines show the limits of the central 68\% of the distribution and the text above the lines gives their values in units pixel. Ranking centre estimators from the most stable to the least stable, we get: $\psi(-0.5)=\psi(0.5)$, the median, the mean, and the mode.}
    \label{fig:centre_comparison_hist}
\end{figure}

The mean values for all four estimators are different and they are not consistent with each other based on the widths of their distributions (see the inset in \figref{fig:centre_comparison_ip_model}). The largest difference between any two estimators is $\approx \pix{0.12}$ (between the median and \ak's estimator) and the smallest difference is $\approx\pix{0.01}$ (between the median and the mean). Distributions for all four estimators show a single peak and are approximately symmetric around their central value (see \figref{fig:centre_comparison_hist}). Concentrating on the distribution width, we see significant differences across the estimators. The text in each of the four panels of the same Figure gives the central 68\% distribution limits, when the distribution median is set to zero. Ranking the four estimators according to their distribution widths (from the most narrow to the most broad), we obtain: the \ak's estimator ($\psi(\Delta x  = -0.5)=\psi(\Delta x = 0.5)$), the median, the mean, and the mode. 

We therefore chose to centre all our empirical IP profiles $\psi$ at the location which results in $\psi(\Delta x  = -0.5)=\psi(\Delta x = 0.5)$. This is also reflected in \figref{fig:centre_comparison_ip_model}, which is plotted with its centre at that location. 
\newpage
\FloatBarrier
\section{A full Markov Chain Monte Carlo approach}\label{app:mcmc}

\begin{figure}
    \centering
    \includegraphics[width=\columnwidth]{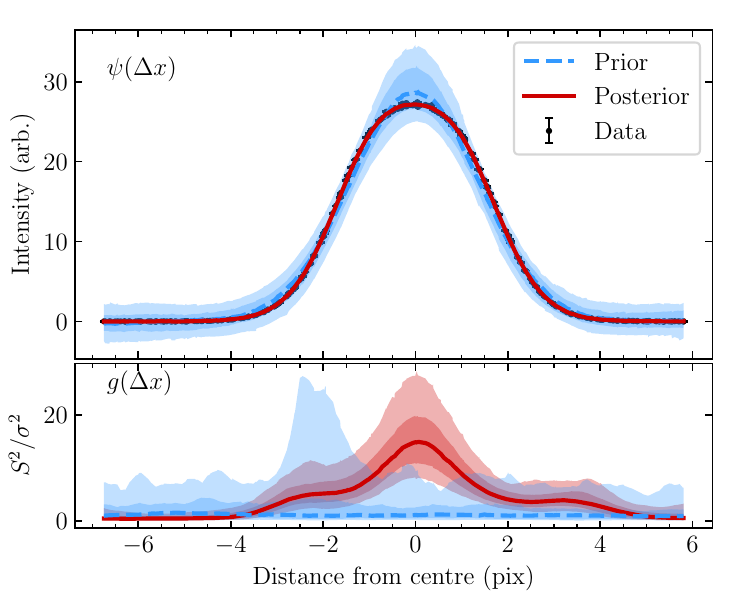}
    \caption{Prior and posterior distributions for $\psi(\Delta x)$ and $g(\Delta x)$. Top panel: The data correspond to the tenth segment of order 110. The dashed blue line is the median of the prior distribution for $\psi(\Delta x)$. The blue shaded bands around the blue line show the central 68\% and 95\% of the prior distribution. The red solid line and the corresponding red bands show the same quantities for the posterior distribution. Bottom panel: The lines and the bands have the same meaning as in the top panel but are for function $g(\Delta x)$ which modifies the variance array as per Eq.~\eqref{eq:epsilon_i_modified}.}
    \label{fig:MCMC_prior_vs_posterior}
\end{figure}

All of the analysis presented in the main body of this work optimises the GP hyperparameters. While our results show that this is sufficient to describe the data, we wanted to check whether uncertainty in the hyperparameter values may significantly affect our conclusions. To test this, we sampled the posterior distrubution of the hyperparameters via Markov Chain Monte Carlo (MCMC) algorithm for several randomly chosen segments. Specifically, we used the Hamiltonian Monte Carlo/No-U-Turn sampler \citep{HMC_NUTS} as implemented in the \texttt{numpyro} \citep{Bingham2019_pyro,Phan2019_numpyro} software. The posterior comprises a likelihood identical to that used in section \secref{sec:empirical_variance} (i.e. with one GP for the line profile and another for the excess variance) along with prior distributions on the hyperparameters. For these `hyperpriors', we used normal distributions, whose mean and variance were set to some reasonable number (see \figref{fig:MCMC_prior_vs_posterior}). We ran four chains using 300 and 600 sampling steps. We assessed convergence of the chains via two diagnostics: the split Gelman-Rubin $\hat{R}$ \citep{Rhat} and effective sample size $N_\mathrm{eff}$ \citep{Geyer_introMCMC}. The values of these diagnostics for all fitted parameters are shown in \tabref{tab:mcmc_statistics}. 

\begin{figure}
    \centering
    \includegraphics[width=\columnwidth]{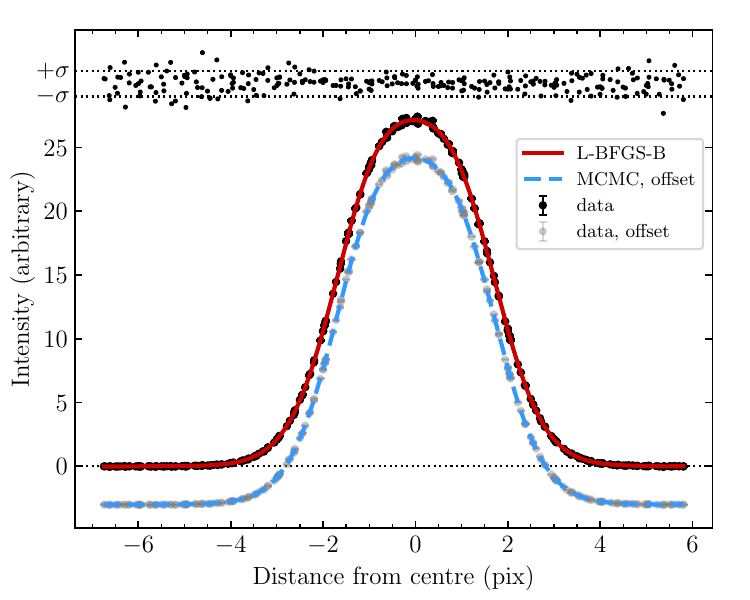}
    \caption{Comparison between IP models derived in two different ways. The model derived using the L-BFGS-B is shown as a solid red line (overlaid on top of black points) and the MCMC derived model is shown as a dashed blue line (overlaid on top of grey points). The two IP models are indistinguishable from one another. The two sets of lines and data points have been offset by 3 in $y$ direction for clarity. The normalised residuals between the two models, that is the MCMC model minus the L-BFGS-B model divided by the error on the data, are shown as a black dots at the top of the panel. The two dotted horizontal lines bracketing the residuals indicate $\pm 1\sigma$, where $\sigma$ is the error on the data. }
    \label{fig:MCMC_vs_LBFGSB}
\end{figure}

The median IP model derived through MCMC is indistinguishable from the IP model derived through the L-BFGS-B method and the differences between the two models are consistent with the noise (\figref{fig:MCMC_vs_LBFGSB}). The posterior probability distributions for hyperparameters $ A, \mu, \sigma, y_0, a,l$ for the instrumental profile, and $(a_g,l_g)$ for the excess variance are shown in \figref{fig:MCMC_parameters}. The hyperparameter values returned by the L-BFGS-B method for the same data are also shown as red lines and squares, and printed out in the top right corner of the figure. Hyperparameter values associated with the IP shape returned by the L-BFGS-B method all fall within the central 68\% of the posterior distribution determined using MCMC. 

\begin{figure}
    \centering
    \includegraphics[width=\columnwidth]{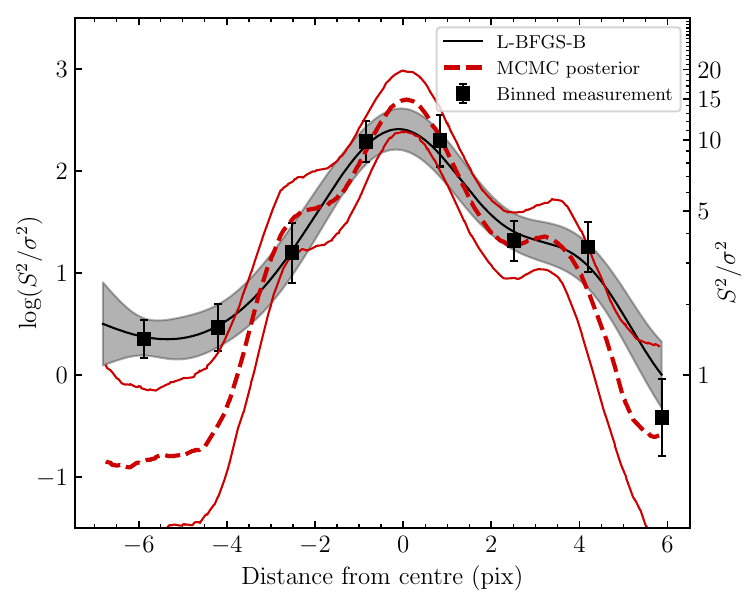}
    \caption{Comparison of $g(\Delta x)$ derived using MCMC and L-BFGS-B. The black squares and the solid black line show the method for variance modification based on L-BFGS-B presented in the main body of the paper and implemented in this work. The dashed red line shows the median of the MCMC posterior distribution for $g(\Delta x)$ and the thin solid red lines show the central 68\% limits. The two models are in excellent agreement.}
    \label{fig:MCMC_vs_LBFGSB_scatter}
\end{figure}

\begin{figure*}
    \centering
    \includegraphics[width=\textwidth]{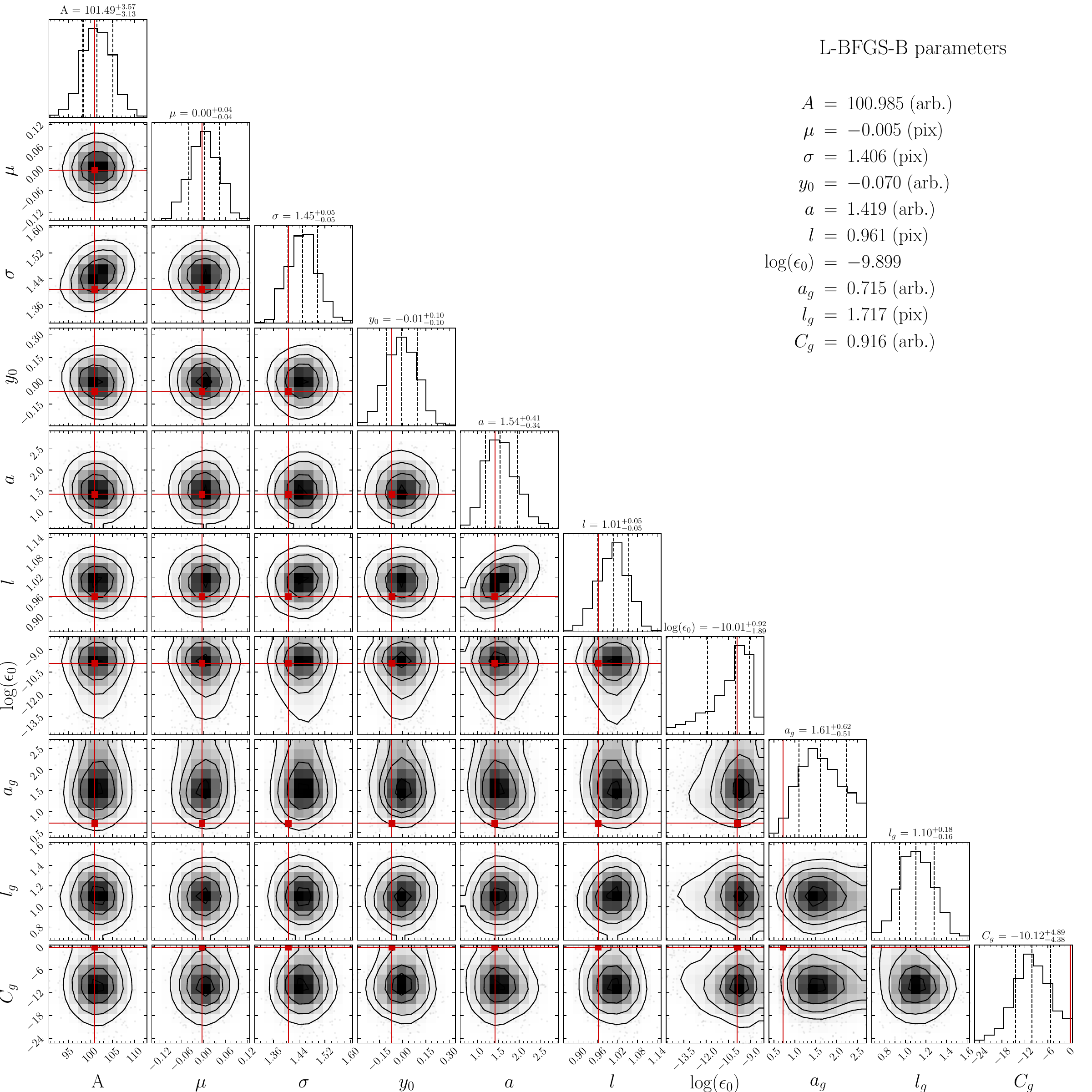}
    \caption{Posterior distribution of hyperparameters in MCMC calculations. The corresponding L-BFGS-B values are printed in the top right corner and are indicated by red lines in all panels.}
    \label{fig:MCMC_parameters}
\end{figure*}

On the other hand, hyperparameters controlling the variance modification all fall outside of the central 68\% of their respective MCMC posterior distributions. However, overplotting the $g(\Delta x)$ from both methods in \figref{fig:MCMC_vs_LBFGSB_scatter}, we see that the two functions agree very well so this disagreement in hyperparameter values is assumed not to be problematic. 

\begin{table}
    \centering
    \caption{The convergence of the MCMC chains was confirmed using two statistics: $\hat{R}$ (should be near 1) and $N_\mathrm{eff}$ (should be over 50). }
    \begin{threeparttable}
    \begin{tabular}{c S[table-format=3.3] S[table-format=4.3]}
         Quantity      &   \multicolumn{1}{c}{$\hat{R}$}     & \multicolumn{1}{c}{$N_{eff}$} \\
         \hline\hline
         $A$                & 1.000  & 1262.332 \\ 
         $\mu$              & 1.001  & 913.540 \\ 
         $\log(\sigma)$     & 1.001  & 1587.845 \\ 
         $y_0$              & 0.999  & 1070.650 \\ 
         $\log(a)$          & 0.999  & 959.304 \\ 
         $\log(l)$          & 1.001  & 810.947 \\ 
         $\log(\sigma_0)$ & 1.014  & 93.589 \\ 
         $\log(a_g)$        & 1.006  & 198.493 \\ 
         $\log(l_g)$        & 0.999  & 54.589 \\ 
         $\log(C_g)$        & 0.999  & 179.070 \\ 
         $\psi(\Delta x)$   & 1.005\tnote{*}  & 711.299\tnote{*} \\ 
         $g(\Delta x)$      & 1.091\tnote{*}  & 82.120\tnote{*} \\ 
    \end{tabular}
    \begin{tablenotes}
        \item [*] Average over the data points.
    \end{tablenotes}
    \end{threeparttable}
    
    \label{tab:mcmc_statistics}
\end{table}

\FloatBarrier
\section{Details of a single example model}\label{app:ip_example}

\begin{figure*}
    \centering
    \includegraphics[width=\textwidth]{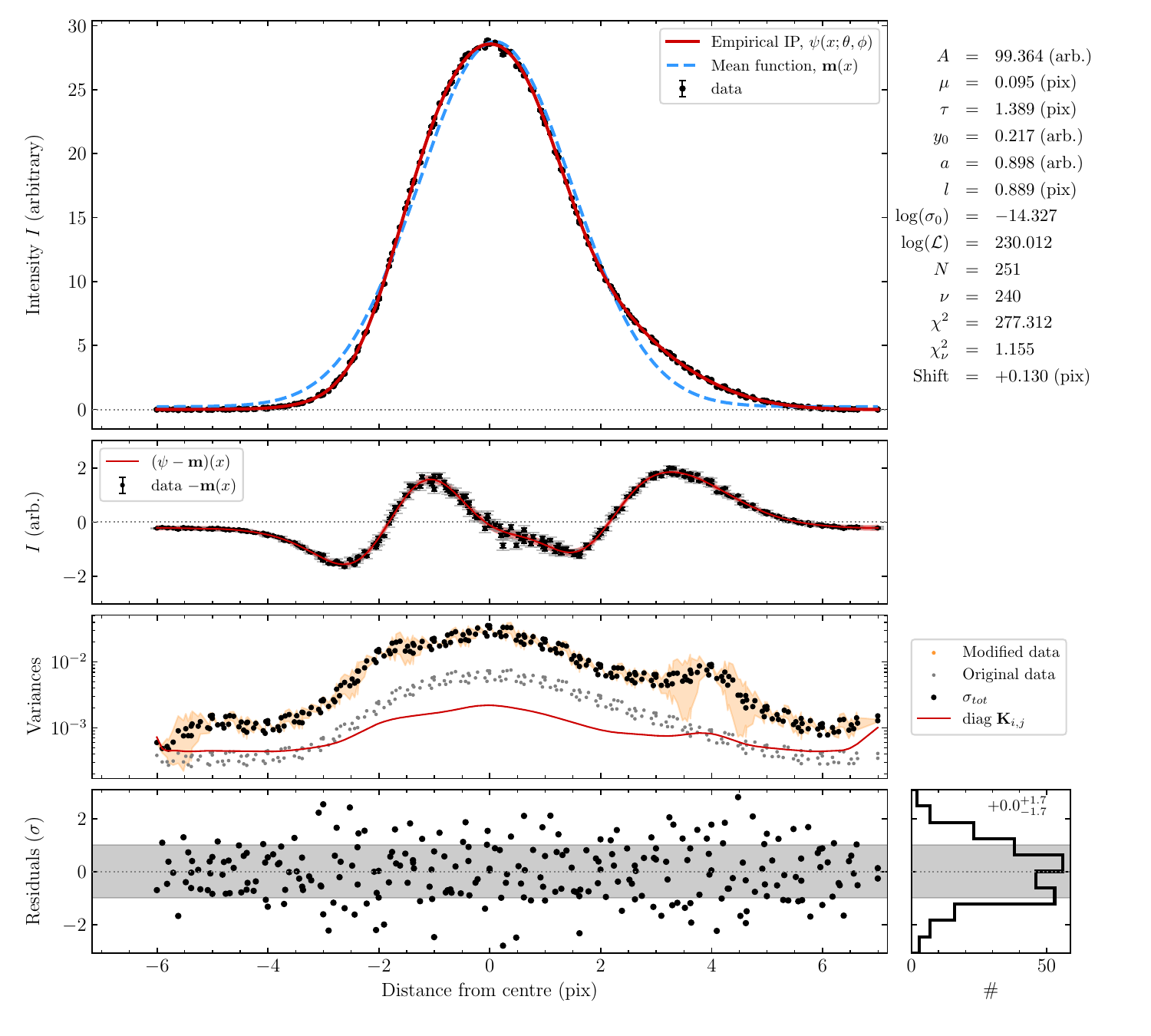}
    \caption{Detailed view of the IP model for a single segment. First panel from the top: The black points are the 251 samples of the IP, $\hat{\psi}$, in the first segment of order 90 derived from flux normalised and stacked astrocomb lines (Eq.~\eqref{eq:effective_IP_estimator}). The thick red line is the empirical IP model, $\psi(x)$ in the first iteration. The model was derived by performing a GP regression on the black points (\secref{sec:GP_regression}) with a secondary GP for variance estimation. The dashed blue line is the mean function of the GP, Eq.~\eqref{eq:gp_mean_function}. The numbers to the right of the panel give additional information. The first seven numbers are the values of the parameters ($\theta,\phi,\sigma_0$) that minimise the loss function, $-\log\mathcal{L}$, where $\mathcal{L}$ is the model likelihood (also quoted). $N$ and $\nu$ are the number of data points and the number of degrees of freedom in the fit, respectively. $\chi^2$ is given by Equation \eqref{eq:chisq} and $\chi^2_\nu=\chi^2/\nu$. The final number specifies the shift applied to $\Delta x$ to ensure proper centring, Eq.~\eqref{eq:anderson_shift}. 
    Second panel: The black points show the difference between the data and the GP mean function (a Gaussian with parameters $\theta=\{A,\mu,\sigma, y_0\}$). The red line shows $\psi(x)-\mathbf{m}(x)$, i.e. departures from the Gaussian shape. 
    Third panel: Grey dots are the variances on the $\hat{\psi}$, i.e.\ Eq.\ \eqref{eq:effective_IP_variance}. Orange points and bands are the variances inferred from the data (as per Eq.~\eqref{eq:epsilon_i_modified}) and the corresponding uncertainty. The larger black dots are the sum of modified $\epsilon_{\rm data}$ and $\sigma_0$. The red line is the variance of the GP model, i.e.\ the diagonal of $\mathbf{K}_{i,j}$, Equation \eqref{eq:gp_correlation_matrix}.
    Fourth panel: The grey dots show the normalised residuals. The grey shaded area shows the $\pm1\sigma$ range and the dashed black lines show the 5th and 95th percentiles. The small panel to the right shows the histogram of the residuals. The shaded area and the dashed lines are the same as in the panel immediately to the left. The number in the top right corner of the panel is the median, and the upper and lower limits correspond to the 5th and 95th percentiles. 
    }
    \label{fig:IP_example_detailed_scatter=True}
\end{figure*}
\newpage
\FloatBarrier
\section{IP models in velocity space}\label{sec:IP_velocity}

\begin{figure*}
    \centering
    \includegraphics[height=0.9\textheight,keepaspectratio]{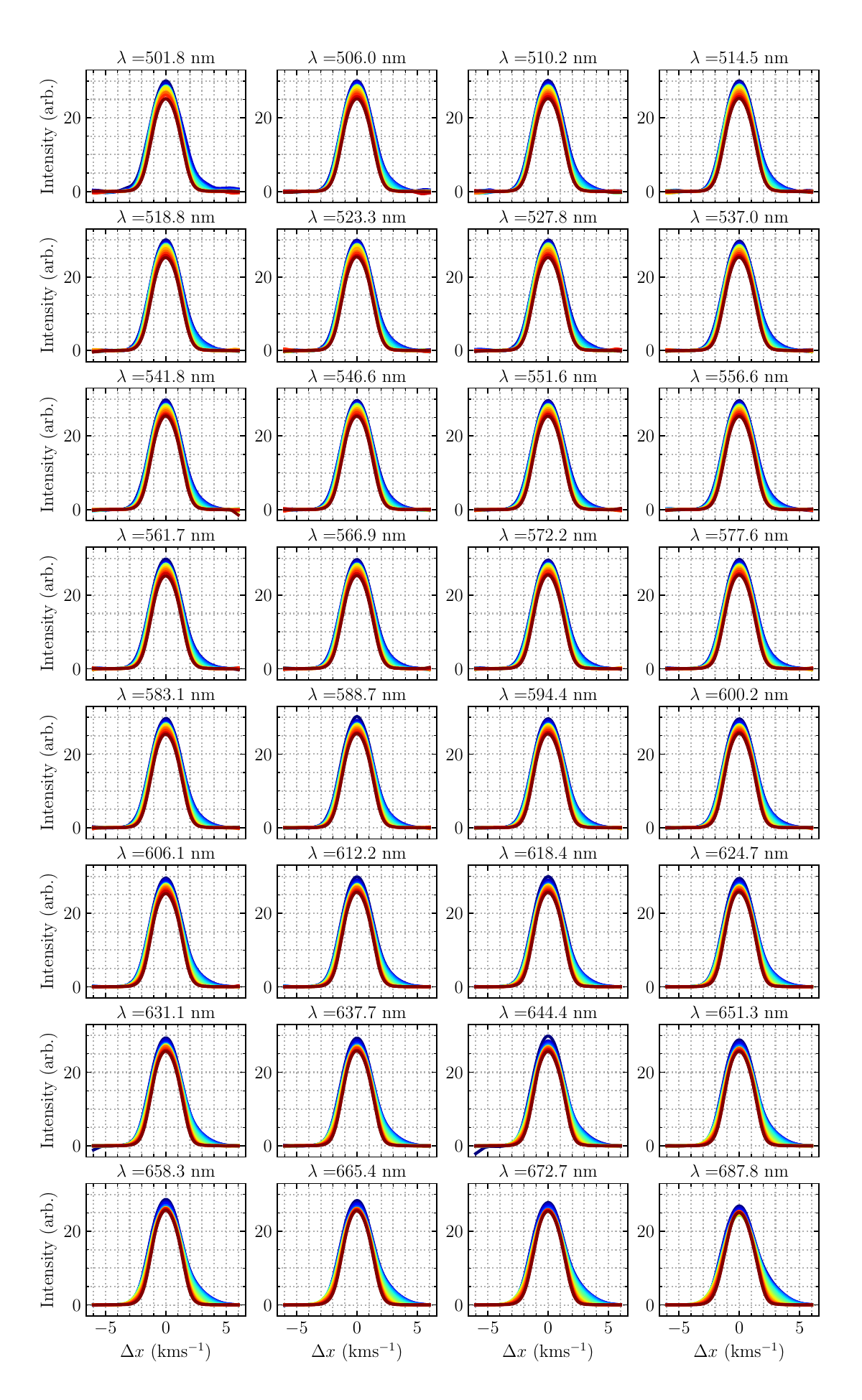}
    \caption{HARPS IP models in velocity space for orders 89 through 122. Each panel shows the IP in 16 segments of one order, whose central wavelength is printed above it. Line colour within the panel changes from blue to red with increasing wavelength. }
    \label{fig:IP_all_velocity}
\end{figure*}

\FloatBarrier
\section{Table with coefficients for intra-order distortions}\label{app:distortion_coefficients}
\begin{table}
\centering
    \caption{Coefficients for a second order polynomial describing intra-order wavelength scale distortions associated with Gaussian IP use, i.e.\ $\epsilon(\lambda, \lambda_0) = \sum_{i=0}^2 c_i(\lambda-\lambda_0)^i$. Order 115 falls in between two CCDs comprising the HARPS detector so is not recorded. 
   }
    
    \begin{tabular}{c S[table-format=3.3] S[table-format=3.2] S[table-format=3.2] S[table-format=3.1] }
        \hline\hline
        \multirow{2}{*}{Order} & $\lambda_{0}$ & $c_{2}$ & $c_{1}$ & $c_{0}$  \\
              & {(\SI{}{\nano\meter})}      & {(\SI{}{\meter\per\second\per\nano\meter\squared})} & {(\SI{}{\meter\per\second\per\nano\meter})} & {(\mps{})} \\
        \hline
  122 & 501.804 & \num{ -0.4(3)} & \num{  1.5(5)} & \num{   -2(1)}\\
  121 & 505.814 & \num{ -0.6(1)} & \num{  2.5(2)} & \num{ -2.4(4)}\\
  120 & 510.027 & \num{-0.65(9)} & \num{  2.6(1)} & \num{ -2.6(3)}\\
  119 & 514.319 & \num{-0.67(9)} & \num{  2.6(1)} & \num{ -2.5(3)}\\
  118 & 518.675 & \num{-0.67(9)} & \num{  2.7(1)} & \num{ -2.6(3)}\\
  117 & 523.098 & \num{ -0.7(1)} & \num{  2.6(2)} & \num{ -2.8(4)}\\
  116 & 527.614 & \num{-0.68(9)} & \num{  2.8(1)} & \num{ -2.9(3)}\\
  114 & 536.865 & \num{-0.68(8)} & \num{  2.6(1)} & \num{ -2.4(3)}\\
  115 & {\ldots} & {\ldots} & {\ldots} & {\ldots}\\
  113 & 541.623 & \num{-0.64(7)} & \num{  2.6(1)} & \num{ -2.6(3)}\\
  112 & 546.447 & \num{-0.70(7)} & \num{  2.7(1)} & \num{ -2.5(3)}\\
  111 & 551.377 & \num{-0.72(7)} & \num{  2.7(1)} & \num{ -2.4(3)}\\
  110 & 556.378 & \num{-0.69(7)} & \num{  2.6(1)} & \num{ -2.6(3)}\\
  109 & 561.489 & \num{-0.67(7)} & \num{  2.6(1)} & \num{ -2.6(3)}\\
  108 & 566.676 & \num{-0.66(6)} & \num{  2.7(1)} & \num{ -2.8(3)}\\
  107 & 571.979 & \num{-0.68(6)} & \num{  2.7(1)} & \num{ -2.6(3)}\\
  106 & 577.362 & \num{-0.68(6)} & \num{  2.6(1)} & \num{ -2.7(3)}\\
  105 & 582.868 & \num{-0.65(6)} & \num{  2.7(1)} & \num{ -2.8(3)}\\
  104 & 588.470 & \num{-0.63(7)} & \num{  2.5(1)} & \num{ -2.9(3)}\\
  103 & 594.180 & \num{-0.67(5)} & \num{ 2.65(9)} & \num{ -3.0(3)}\\
  102 & 600.002 & \num{-0.70(6)} & \num{  2.6(1)} & \num{ -2.9(3)}\\
  101 & 605.940 & \num{-0.66(6)} & \num{  2.7(1)} & \num{ -3.1(3)}\\
  100 & 611.996 & \num{-0.68(6)} & \num{  2.9(1)} & \num{ -3.2(3)}\\
  99 & 618.174 & \num{-0.69(6)} & \num{  2.9(1)} & \num{ -3.5(3)}\\
  98 & 624.490 & \num{-0.70(5)} & \num{  3.0(1)} & \num{ -3.6(3)}\\
  97 & 630.913 & \num{-0.75(6)} & \num{  3.1(1)} & \num{ -3.6(3)}\\
  96 & 637.493 & \num{-0.80(6)} & \num{  3.2(1)} & \num{ -3.8(3)}\\
  95 & 644.225 & \num{-0.87(7)} & \num{  3.3(1)} & \num{ -3.7(4)}\\
  94 & 651.049 & \num{-0.86(7)} & \num{  3.7(1)} & \num{ -4.3(4)}\\
  93 & 658.046 & \num{-0.91(6)} & \num{  3.8(1)} & \num{ -4.6(4)}\\
  92 & 665.208 & \num{-1.02(6)} & \num{  4.2(1)} & \num{ -4.9(4)}\\
  91 & 672.500 & \num{-1.11(6)} & \num{  4.6(1)} & \num{ -5.3(4)}\\
  90 & 679.982 & \num{-1.20(7)} & \num{  4.8(1)} & \num{ -5.8(5)}\\
  89 & 687.629 & \num{-1.42(7)} & \num{  5.4(1)} & \num{ -5.8(5)}\\
  \hline
    \end{tabular}
       \vspace{0.4cm}
    
    \label{tab:distortion_coefficients}
\end{table}
\end{appendix}
\end{document}